\documentclass[prr,aps,amsmath,amssymb,superscriptaddress,twocolumn]{revtex4-1}
\usepackage[pdftex]{graphicx}

\usepackage{bm}
\usepackage{amsmath}
\usepackage{color}
\usepackage{setspace}
\usepackage{amssymb}
\usepackage{latexsym}
\usepackage{mathtools}
\usepackage{lipsum}
\usepackage{braket}

\begin{document}
	\title{Spin-wave growth via Shapiro resonances in a spinor Bose-Einstein condensate} 
	
	\author{Yuya Imaeda}
	\affiliation{Department of Applied Physics, Nagoya University, Nagoya 464-8603, Japan}
	
	\author{Kazuya Fujimoto}
	\affiliation{Department of Applied Physics, Nagoya University, Nagoya 464-8603, Japan}
	\affiliation{Institute for Advanced Research, Nagoya University, Nagoya 464-8601, Japan}
	
	\author{Yuki Kawaguchi}
	\affiliation{Department of Applied Physics, Nagoya University, Nagoya 464-8603, Japan}	
	
	\begin{abstract}
		We theoretically study the resonant phenomenon in a spin-1 Bose-Einstein condensate periodically driven by a quadratic Zeeman coupling. This phenomenon is closely related to the Shapiro steps in superconducting Josephson junctions, and the previous experimental work [Evrard $et~al.,$ Phys. Rev. A $\bm{100}$, 023604 (2019)] for a spin-1 bosonic system observed the resonant dynamics and then called it Shapiro resonance. In this work, using the spin-1 Gross-Pitaevskii equation, we study the Shapiro resonance beyond the single-mode approximation used in the previous work, which assumes that all components of the spinor wavefunction have the same spatial configuration. Considering resonant dynamics starting from a polar state, we analytically calculate the Floquet-Lyapunov exponents featuring an onset of the resonance under a linear analysis and find that spin waves with finite wavenumbers can be excited. This kind of non-uniform excitation cannot be described by the single-mode approximation. Furthermore, to study the long-time resonant dynamics beyond the linear analysis, we numerically solve the one-dimensional spin-1 Gross-Pitaevskii equation, finding that the nonresonant hydrodynamic variables also grow at wavelengths of even multiples of the resonant one due to the nonlinear effect.
	\end{abstract}

	\date{\today}
	\maketitle
	
    \section{Introduction}
	The engineering of quantum systems by periodic driving has drawn great attention over a decade, and ultracold atoms have become a promising platform for realizing such driven quantum systems due to their high experimental controllability~\cite{Floquet1,Floquet2,Floquet3}. Indeed, applying various periodic modulations to ultracold atoms, recent experiments have realized several topological models such as the Haldane model and the Hofstadter-Harper model \cite{topo1,topo2,topo3}, and have also observed exotic phases of matters such as a time crystal~\cite{TC1,TC2}.
	
	Such engineering by external driving is recently utilized to generate quantum entanglement in a spin-1 Bose-Einstein condensate (BEC) \cite{Shapiro2}, which is comprised of spin-1 bosons characterized by the three magnetic sublevels $m=1,0$, and $-1$~\cite{YK2012,MU2013}. Modulating a quadratic Zeeman (QZ) coupling by microwaves, Evrard $et~al.$~\cite{Shapiro2} induce resonance between the different magnetic sublevels and observe entangled spin states. This resonant phenomenon is essentially the same as the Shapiro steps originally investigated in Josephson junctions between superconductors~\cite{Shapiro_org} and thus is called Shapiro resonance. In the series of these works \cite{Shapiro1,Shapiro2}, they have investigated the Shapiro resonance mainly in a restricted situation where a single-mode approximation is valid. Under this approximation, all three magnetic sublevels have the same spatial configuration, and thus we cannot address spin-wave excitation with a finite wavelength as well as nontrivial magnetic pattern formations~\cite{Stamper1,Stamper2}. 
	
	In this work, we theoretically study the Shapiro resonance in a spin-1 BEC beyond the single-mode approximation. Using the Gross-Pitaevskii equation (GPE)~\cite{SGP1,SGP2}, we investigate the resonant dynamics starting from a polar state by periodically modulating the QZ coupling. First, we analytically derive the resonant condition by linearizing the spin hydrodynamic equations equivalent to the GPE~\cite{spinhydro} and subsequently applying the Floquet's theorem to the linearized equations. Here, the resonant condition is identified by a nonzero Floquet-Lyapunov (FL) exponent, which captures an onset of the resonance. Second, we numerically solve the one-dimensional (1D) GPE, demonstrating the validity of our linear analysis and investigating the nonlinear dynamics over a long time. Spin-wave excitations due to the nonlinear effects are explained from the constraints on the hydrodynamic variables. Finally, we discuss the experimental possibilities by using the parameters used in the previous experiments~\cite{Rb_Hirano,Na_Shin,Na_Raman,Na_Dalibard,Li_Choi}.
	
	The rest of this article is organized as follows. In Sec.~\ref{sec:models}, we introduce the GPE and the spin hydrodynamic equations for a spin-1 spinor BEC. In Sec.~\ref{sec:linear}, we derive the analytical expressions for the resonant conditions by a linear analysis of the spin hydrodynamic equations. In Sec.~\ref{Sec:num}, we show the numerical results of the 1D GPE, showing that our analytical results work well and investigating the long-time resonant dynamics. In Sec.~\ref{Sec:discussion}, we discuss the experimental possibilities for observing non-uniform Shapiro resonances. 
	The summary for this work is given in Sec.~\ref{Sec:conclusion}.

	\section{Models \label{sec:models}}
	In this work, we study the dynamics of a spin-1 spinor BEC within the mean-field approximation. Here, we introduce the mean field equation, namely the GPE, and subsequently explain the spin hydrodynamic form~\cite{spinhydro}, which is quite convenient for analytically investigating the Shapiro resonance in Secs.~\ref{sec:linear} and \ref{Sec:num}.
    
	\subsection{Spin-1 Gross-Pitaevskii equation}
	We consider a spin-1 BEC in a uniform system without a trapping potential. Under the mean-field approximation, the macroscopic wave functions $\psi_m$ of atoms in the magnetic sublevel $m=1,0,-1$ obey the following GPE:
	\begin{align}
		\nonumber i\hbar\frac{\partial}{\partial t}\psi_m=&\left(-\frac{\hbar^2}{2M}\nabla^2+q(t)m^2\right)\psi_m
		\\&+c_0\rho\psi_m+c_1\sum_{m'=0,\pm1}\bm{F}\cdot\hat{\bm{\mathrm{f}}}_{mm'}\psi_{m'}, \label{eq:GP3D}
	\end{align}
	where $M$ is the atomic mass, and $c_0$ and $c_1$ are the strength of the spin-independent and spin-dependent interaction, respectively. The sign of $c_1$ determines the magnetism of the system: The condensate is antiferromagnetic (AFM) for $c_1>0$ and ferromagnetic (FM) for $c_1<0$. The total particle number density $\rho$ and the spin density $F_\mu~(\mu=x, y, z)$ are defined by
	\begin{align}
		\rho&=\sum_{m=0,\pm1}|\psi_m|^2, \label{eq:def_rho}
		\\F_\mu&=\sum_{m,m'=0,\pm1}\psi_m^*(\hat{\mathrm{f}}_\mu)_{mm'}\psi_{m'},
	\end{align}
	where $\hat{\mathrm{f}}_\mu~(\mu=x,y,z)$ is the $\mu$-component of the spin-1 matrix given by
	\begin{align*}
	    \hat{\mathrm{f}}_x&=\frac{1}{\sqrt{2}}
	    \begin{pmatrix} 0 & 1 & 0 \\ 1 & 0 & 1 \\ 0 & 1 & 0 \end{pmatrix},\ \ 
	    \hat{\mathrm{f}}_y=\frac{1}{\sqrt{2}}
	    \begin{pmatrix} 0 & -i & 0 \\ i & 0 & -i \\ 0 & i & 0 \end{pmatrix},\\
	    \hat{\mathrm{f}}_z&=
	    \begin{pmatrix} 1 & 0 & 0 \\ 0 & 0 & 0 \\ 0 & 0 & -1 \end{pmatrix}.
	\end{align*}
	We assume that the QZ coupling strength $q(t)$ consists of a static term and an oscillating term with frequency $\Omega$ as given by
	\begin{align}
		q(t)=q_0+q_\mathrm{osc}\sin(\Omega t),
	\end{align}
	where $q_0$ and $q_{\rm osc}$ are the strength of the static and oscillating QZ couplings. 
	Here, we eliminate the linear Zeeman coupling without loss of generality because it can be removed when we move onto the rotating frame of reference in spin space with the Larmor frequency~\cite{YK2012}.

	\subsection{Spin hydrodynamic equations}
	We can rewrite GPE~\eqref{eq:GP3D} without any approximation to the equations of motion for the particle number density $\rho$, the particle density current $v_\mu$, the spin density vector $f_\mu$, and the nematic density tensor $n_{\mu\nu}$ $(\mu, \nu=x, y, z)$, which are defined by Eq.~\eqref{eq:def_rho} and
	\begin{align}
	v_\mu&=\frac{\hbar}{2Mi}\sum_{m=0,\pm1}\left[ \zeta_m^*(\nabla_\mu\zeta_m)-(\nabla_\mu \zeta_m^*)\zeta_m \right],
	\\f_\mu&=\sum_{m,m'=0,\pm1}\zeta_m^*(\hat{\mathrm{f}}_\mu)_{mm'}\zeta_{m'},
	\\n_{\mu\nu}&=\sum_{m,m'=0,\pm1}\zeta_m^*\left( \frac{\hat{\mathrm{f}}_\mu \hat{\mathrm{f}}_\nu+\hat{\mathrm{f}}_\nu \hat{\mathrm{f}}_\mu}{2} \right)_{mm'}\zeta_{m'},
	\end{align}
	where $\zeta_m\equiv\psi_m/\sqrt{\rho}$ is the normalized spinor wavefunction. 
	By taking the time derivative of the above quantities and using GPE~\eqref{eq:GP3D},
	we obtain the following spin hydrodynamic equations~\cite{spinhydro}:
	\begin{align}
	\label{eq:SHE1}
	\frac{\partial \rho}{\partial t}&+\bm{\nabla}\cdot(\rho\bm{v})=0,
	\\\label{eq:SHE2}
	\frac{\partial(\rho f_\mu)}{\partial t}&+\bm{\nabla}\cdot\left(\rho\bm{v}^{{\rm (s)}}_\mu\right)=-\frac{2}{\hbar}\rho q\sum_{\nu=x,y,z}\epsilon_{z\mu\nu}n_{z\nu},
	\\\nonumber\frac{\partial (\rho n_{\mu\nu})}{\partial t}&+\bm{\nabla}\cdot\left(\rho\bm{v}^{{\rm (n)}}_{\mu\nu}\right)
	\\\nonumber&=-\frac{1}{2\hbar}\rho q \sum_{\lambda=x,y,z}(\epsilon_{z\mu\lambda}\delta_{z\nu}f_\lambda+\epsilon_{z\nu\lambda}\delta_{z\mu}f_\lambda)
	\\\label{eq:SHE3}&+\frac{c_1}{\hbar}\rho^2\sum_{\lambda,\eta=x,y,z}(\epsilon_{\mu\lambda\eta}f_\lambda n_{\nu\eta}+\epsilon_{\nu\lambda\eta}f_\lambda n_{\mu\eta}),
	\end{align}
	\begin{align}
	\nonumber\frac{\partial v_\mu}{\partial t}&+\sum_{\nu=x,y,z}(v_\nu\nabla_\nu)v_\mu-\frac{\hbar^2}{2M^2}\nabla_\mu\frac{\bm{\nabla}^2\sqrt{\rho}}{\sqrt{\rho}}
	\\\nonumber&+\frac{\hbar^2}{4M^2\rho}\sum_{\nu,\lambda=x,y,z}\nabla_\nu\rho\Bigl\{ \frac{1}{2}\left[ (\nabla_\mu f_\lambda)(\nabla_\nu f_\lambda)-f_\lambda(\nabla_\mu\nabla_\nu f_\lambda) \right]
	\\\nonumber&+\sum_{\eta=x,y,z}\left[ (\nabla_\mu n_{\lambda\eta})(\nabla_\nu n_{\lambda\eta})-n_{\lambda\eta}(\nabla_\mu\nabla_\nu n_{\lambda\eta}) \right] \Bigr\}
	\\&\label{eq:SHE4}=-\frac{1}{M}\{c_0\nabla_\mu\rho+c_1\sum_{\nu=x,y,z}f_\nu(\nabla_\mu\rho f_\nu) \}.
	\end{align}
	Here, we have defined the spin current $\bm{v}_\mu^{\rm (s)}$ and the nematic current $\bm{v}_{\mu\nu}^{\rm (s)}$ as	
	\begin{align}
	\nonumber\bm{v}_\mu^{\rm (s)}&= f_\mu\bm{v}-\frac{\hbar}{M}\sum_{\nu\lambda=x,y,z}\epsilon_{\mu\nu\lambda}\Bigl[ \frac{1}{4}f_\nu(\bm{\nabla} f_\lambda)
	\\&\ \ \ \ \ \ \ \ \ \ \ \ \ \ \ +\sum_{\eta=x,y,z}n_{\nu\eta}(\bm{\nabla}n_{\lambda\eta}) \Bigr],
	\\\nonumber\bm{v}_{\mu\nu}^{\rm (n)}&= n_{\mu\nu}\bm{v}-\frac{\hbar}{4M}\sum_{\lambda,\eta=x,y,z}\bigg\{ \epsilon_{\mu\lambda\eta}\left[f_\lambda(\bm{\nabla }n_{\nu\eta})-(\bm{\nabla}f_\lambda)n_{\nu\eta}\right]
	\\&\ \ \ \ \ \ \ \ \ \ \ \ \ \ \ +\epsilon_{\nu\lambda\eta}\left[ f_\lambda(\bm{\nabla}n_{\mu\eta})-(\bm{\nabla}f_\lambda)n_{\mu\eta} \right] \bigg\}.
	\end{align}
	Note that the number of variables in the spin hydrodynamic equations is larger than in the three-component GPE~\eqref{eq:GP3D}. This is because some of the variables in the spin hydrodynamic equations are dependent on each other via the following constraints~\cite{spinhydro}:
	\begin{align}
	\label{bc:n}\sum_{\mu=x,y,z}n_{\mu\mu}&=2,
	\\\label{bc:nf}\sum_{\nu=x,y,z}n_{\mu\nu}f_\nu&=f_\mu,
	\\\label{bc:det}\frac{1}{4}\sum_{\mu=x,y,z}f_\mu^2&=\det n_{\mu\nu},
	\end{align}
	The detail is described in Ref.~\cite{spinhydro}.
	
	The previous works on Shapiro resonance~\cite{Hoang2016,Shapiro1,Shapiro2} analyze the Madelung form of GPE~\eqref{eq:GP3D}, which is the equations of motion for the density and phase of each component (see Appendix \ref{app2_eqs}), under the single-mode approximation. In this paper, we use the spin hydrodynamic equations rather than the Madelung form since the linear analysis beyond the single-mode approximation becomes simpler for the former case as discussed in Sec.~\ref{sec:linear}.

	\subsection{Parameter setup \label{sec:parameterSetup}}
	\label{sec:params}
    In this study, we prepare a polar BEC, where all atoms are spatially uniform and condensed in the $m=0$ state, and investigate how the numbers of the atoms in the $m=\pm 1$ state increase via parametric resonance. We therefore choose parameters such that the polar state is stable when the oscillating frequency $\Omega$ is off-resonance. This condition is satisfied when $q(t)$ moves in the polar phase region:
	\begin{align}
		\label{ieq:polar}q(t)>\left\{\begin{array}{ll}
		2|c_1|\bar{\rho}&\ (c_1<0);
		\\0& \ (c_1>0),
		\end{array}\right.
	\end{align}
	where $\bar{\rho}$ is the mean particle number density of the condensate. In the following calculations, we choose $q_0,q_{\rm osc}>0$ and $q_0-q_{\rm osc}>{\rm max}(0,-2c_1\bar{\rho})$. We also assume $\Omega>0$ without loss of generality.
	
	The bulk chemical potential in the polar state is given by $c_0\bar{\rho}$.
	The corresponding length and time are given by
	\begin{align}
	    \xi&=\frac{\hbar}{\sqrt{2Mc_0\bar{\rho}}},\\
	    \tau&=\frac{\hbar}{c_0\bar{\rho}},
	\end{align}
	which we use as the characteristic scales of the system.
	
	\section{Linear analysis \label{sec:linear}}
	In this section, we apply linear analysis to the spin hydrodynamic equations around the polar state. We employ the Floquet's theorem to obtain the resonant conditions and the FL exponents analytically.

	\subsection{Linearized spin hydrodynamic equations}
	We discuss the linear stability of the polar state, $(\psi_1,\psi_0,\psi_{-1})=(0,\sqrt{\bar{\rho}},0)$. In the absence of fluctuations, the particle current, the spin density vector, and the nematic density tensor for the polar state are given by
    \begin{align}
    \bar{\bm v}&=\bm 0,\\
    \bar{\bm f}&=\bm 0,\\
	\bar{n}_{\mu\nu}&=\left\{\begin{array}{ll}
	1& \ \ ~(\mu,\nu)=(x,x),(y,y);
	\\0& \ \ ~\mathrm{otherwise}.
	\end{array}\right.
	\end{align}
	We introduce small fluctuations, $\delta\rho,\delta f_\mu, \delta n_{\mu\nu}$, and $\delta v_\mu$, and write the hydrodynamic variables as
	\begin{align}
	\label{eq:fluct1}\rho(\bm{r},t)&=\bar{\rho}+\delta\rho(\bm{r},t),
	\\f_\mu(\bm{r},t)&=\delta f_\mu(\bm{r},t),
	\\n_{\mu\nu}(\bm{r},t)&=\bar{n}_{\mu\nu}+\delta n_{\mu\nu}(\bm{r},t),
	\\\label{eq:fluct4}v_\mu(\bm{r},t)&=\delta v_\mu(\bm{r},t).
	\end{align}
	By substituting them into Eqs.~\eqref{eq:SHE1}--\eqref{eq:SHE4} and expanding the equations up to the first order in the fluctuations, we obtain 13 linearized equations. Among them, the equations for $\delta f_x, \delta f_y, \delta n_{xz}$, and $\delta n_{yz}$ include the driving QZ term and are divided into two sets of coupled equations:
	\begin{align}
	\label{eq:ABx}\hbar\frac{\partial}{\partial t}\left( \begin{array}{c}
	A(\bm{r},t)
	\\B(\bm{r},t)
	\end{array} \right)=\pm\hat{P}(t)\left( \begin{array}{c}
	A(\bm{r},t)
	\\B(\bm{r},t)
	\end{array} \right) ,
	\end{align}
	\begin{align}
	\hat{P}(t)=\left( \begin{array}{cc}
	0 & \displaystyle{ \frac{\hbar^2}{M}\nabla^2-2q(t) }\\
	\displaystyle{ -\frac{\hbar^2}{4M}\nabla^2+\frac{1}{2}q(t)+c_1\bar{\rho} }&0
	\end{array} \right) ,
	\end{align}
	where $(A, B) = (\delta f_x, \delta n_{yz})$ and $(\delta f_y, \delta n_{xz})$. Here, the plus and minus signs on the right-hand side of Eq.~\eqref{eq:ABx} are for $(A, B) =(\delta f_x, \delta n_{yz})$ and $(\delta f_y, \delta n_{xz})$, respectively. These equations indicate that the transverse spin components, $\delta f_x$ and $\delta f_y$, and the off-diagonal elements of the nematic density tensors, $\delta n_{yz}$ and $\delta n_{xz}$, can be amplified by the driving QZ term. Thus, we call them the resonant variables. Note that Eq.~\eqref{eq:ABx} is valid only for the initial stages of the resonant dynamics since we have neglected the nonlinear terms.
	
	\subsection{Application of the Floquet's theorem to Eq.~(\ref{eq:ABx})}
	Equation~$(\ref{eq:ABx})$ can be transformed into an eigenvalue problem of an infinite-dimensional matrix by utilizing the Floquet's theorem, which enables us to derive the resonant conditions. First, we introduce the Fourier transform
	\begin{align}
		\left(\begin{array}{c}
		\tilde{A}(\bm{k},t)\\
		\tilde{B}(\bm{k},t)
		\end{array}\right)&=\int d^dr\left(\begin{array}{c}
		A(\bm{r},t)\\
		B(\bm{r},t)
		\end{array}\right)e^{-i\bm{k}\cdot\bm{r}},
	\end{align}
	and rewrite Eq.~\eqref{eq:ABx} as
	\begin{align}
		\label{eq:AB_FT}\hbar\frac{\partial}{\partial t}\left( \begin{array}{c}
		\tilde{A}
		\\\tilde{B}
		\end{array} \right)=\pm\left( \begin{array}{cc}
		0&-2\epsilon_k-2q(t)\\
		\displaystyle{ \frac{1}{2}\epsilon_k+\frac{1}{2}q(t)+c_1\bar{\rho} }&0
		\end{array} \right)\left( \begin{array}{c}
		\tilde{A}
		\\\tilde{B}
		\end{array} \right),
	\end{align}
	where $\epsilon_k= \hbar^2k^2/(2M)$. Second, we perform the linear transformation such that the time-independent part of the matrix in Eq.~\eqref{eq:AB_FT} is diagonalized. The linear transformation is given by
	\begin{align}
		\left(\begin{array}{c}
		S_A(\bm{k},t)\\
		S_B(\bm{k},t)
		\end{array}\right)&=\ \left(\begin{array}{cc}
		2(\epsilon_k+q_0)&2(\epsilon_k+q_0)\\
		-iE_k&iE_k
		\end{array}\right)^{-1}\left(\begin{array}{c}
		\tilde{A}(\bm{k},t)\\
		\tilde{B}(\bm{k},t)
		\end{array}\right),
	\end{align}
	where $\pm iE_k$ are the eigenvalues of the matrix in Eq.~\eqref{eq:AB_FT} with $q_{\rm osc} = 0$:
	\begin{align}
	    E_k=\sqrt{(\epsilon_k+q_0)(\epsilon_k+q_0+2c_1\bar{\rho})}.
	\end{align}
	Then, Eq.~(\ref{eq:AB_FT}) reduces to
	\begin{align}
		\label{eq:S}\frac{d}{dt}\left( \begin{array}{c}
		S_A(\bm{k},t)\\
		S_B(\bm{k},t)
		\end{array} \right)=\pm\left( \hat{F}+ \frac{q_\mathrm{osc}}{E_k}\sin(\Omega t)\hat{G} \right)\left(\begin{array}{c}
		S_A(\bm{k},t)\\
		S_B(\bm{k},t)
		\end{array}\right)
	\end{align}
	with
	\begin{align}
		\hat{F}&=\frac{i}{\hbar}\left(\begin{array}{cc}
		E_k&0\\
		0&-E_k
		\end{array}\right),
		\\\hat{G}&=\frac{i}{\hbar}\left(\begin{array}{cc}
		\epsilon_k+q_0+c_1\bar{\rho}&-c_1\bar{\rho}\\
		c_1\bar{\rho}&-(\epsilon_k+q_0+c_1\bar{\rho})
		\end{array}\right).
	\end{align}
	Note that $E_k$ is identical to the spin-wave excitation energy obtained by the Bogoliubov analysis under the static QZ term. Equation~(\ref{eq:S}) indeed reproduces the equation of motion for the spin wave when $q_\mathrm{osc}=0$.
	
	Because Eq.~\eqref{eq:S} has the periodicity of $T=2\pi/\Omega$,
	we can apply the Floquet's theorem: Without loss of generality, the functions $S_A$ and $S_B$ can be expanded as
	\begin{align}
		\label{eq:S_expan}\left(\begin{array}{c}
		S_A\\
		S_B
		\end{array}\right)=&e^{\lambda_\mathrm{F}t}\sum_{j=-\infty}^{\infty}\left(\begin{array}{c}
		C_{j,A}\\
		C_{j,B}
		\end{array}\right)\displaystyle{  e^{i\frac{2\pi j}{T}t} },
	\end{align}
	where $\lambda_\mathrm{F}$ is the Floquet exponent. The summation over $j\in \mathbb{Z}$ comes from the Fourier expansion of $T$-periodic functions, and $C_{j,A}$ and $C_{j,B}$ are the Fourier components. By substituting Eq.~\eqref{eq:S_expan} into Eq.~\eqref{eq:S}, we obtain
	\begin{align}
		\label{eq:EigenEq}\hat{Q}\ket{C}=i\lambda_\mathrm{F}\ket{C},
	\end{align}
	where $\hat{Q}$ and $\ket{C}$ are the infinite-dimensional matrix and the column vector, respectively, given by
	\begin{align}
		\label{matrix:Q}\hat{Q}&=\left(\begin{array}{ccccc}
		\ddots&\ddots&&&\mbox{\Huge 0}\\
		\ddots&\hat{F}_{j-1}&-b\hat{G}&&\\
		&b\hat{G}&\hat{F}_j&-b\hat{G}&\\
		&&b\hat{G}&\hat{F}_{j+1}&\ddots\\
		\mbox{\Huge 0}&&&\ddots&\ddots
		\end{array}\right),
		\\
		\ket{C}&=\left(\ldots, C_{j-1,B},\ C_{j,A},\ C_{j,B},\ C_{j+1,A},\ldots\right)^\mathrm{T}
	\end{align}
	with
	\begin{align}
		\hat{F}_j&=\left(\begin{array}{cc}
		j\Omega-E_k/\hbar&0\\
		0&j\Omega+E_k/\hbar
		\end{array}\right),
		\\b&=\frac{q_\mathrm{osc}}{2E_k}.
	\end{align}
	Here, we omit the sign $\pm$ from Eq.~$(\ref{eq:EigenEq})$ because both equations provide the same result about the real part of the Floquet exponents (see Appendix~\ref{app3_Q}). 
    The FL exponent is defined as
	the real part of $\lambda_\mathrm{F}$:
	\begin{align}
		\lambda_\mathrm{FL}=\mathrm{Re}[\lambda_\mathrm{F}].
	\end{align}
    If there exists a positive $\lambda_\mathrm{FL}$, $S_A$ and $S_B$ exhibit exponential increase, that is, the Shapiro resonance. Therefore, we refer to the positive ones as the FL exponents in the following discussions unless otherwise noted.

    In the following sections, we solve the infinite-dimensional eigenvalue problem of Eq.~\eqref{eq:EigenEq} by assuming $b\ll 1$. This assumption is satisfied for large $E_k$. Furthermore, with the parameters discussed in Sec.~\ref{sec:params}, the off-diagonal elements of $\hat{Q}$ is proved to be smaller than $E_k/2$, i.e., $b(\hat{G})_{l,m}<E_k/2$ with $l,m=1,2$, which also supports the validity of the perturbative expansion below. The proof is given in Appendix~\ref{app4_b}.
		
	\begin{figure}[t]
		\begin{center}
			\includegraphics[keepaspectratio, width=8.5cm,clip]{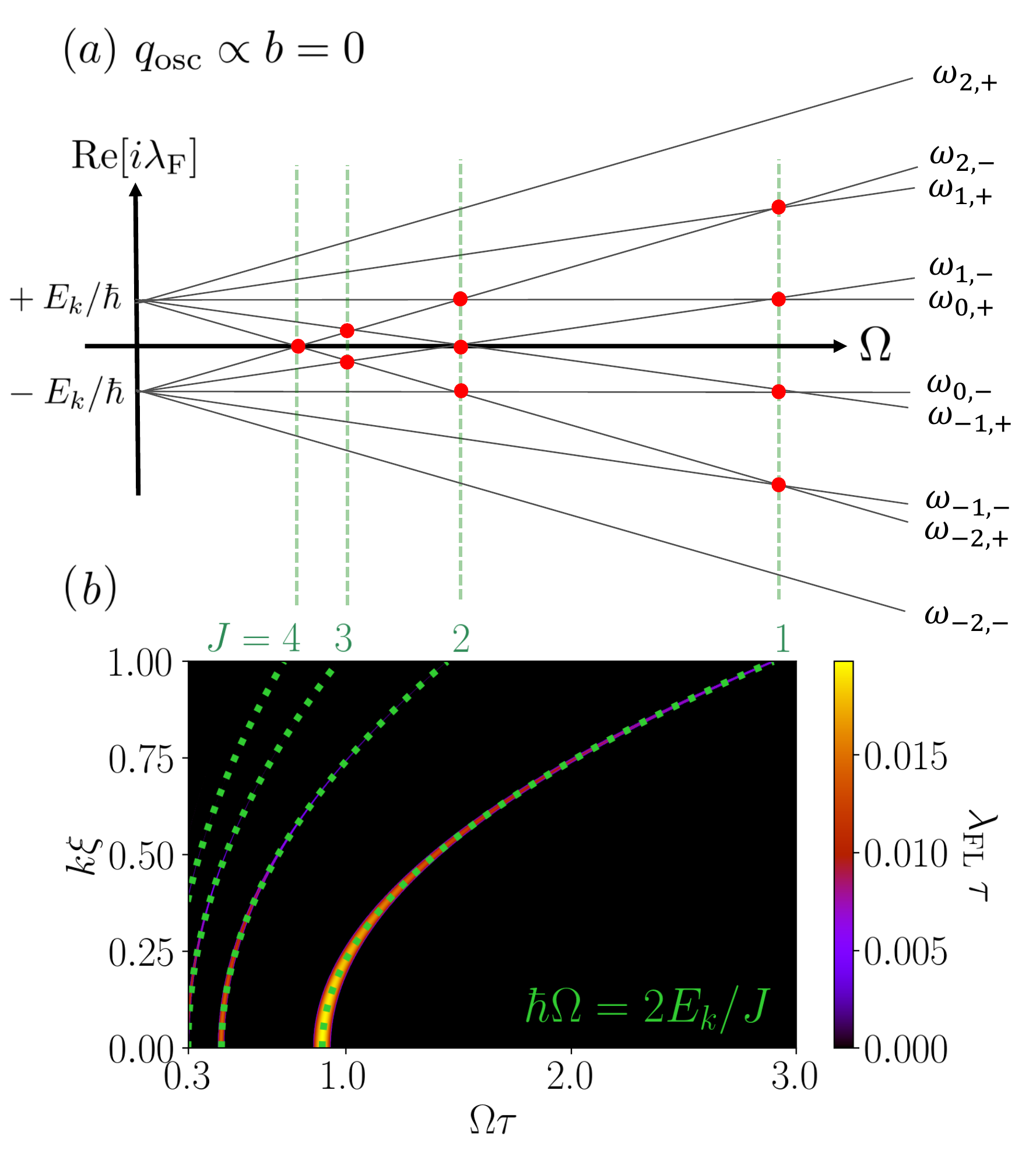}
			\caption{
			(a)~Floquet exponents $\lambda_\textrm{F}$ as functions of driving frequency $\Omega$ at $b=0$, where all $\lambda_\textrm{F}$'s are pure imaginary as given by Eq.~\eqref{eq:lambda_F_b=0}. Shown are  $i\lambda_\mathrm{F}=\omega_{j,\pm}$ with $j=0,\pm1, \pm2$, whose intersection points are denoted by filled red circles. Note that these points are just a subset of the infinite number of the intersection points of $\lambda_\textrm{F}$'s. 
			(b)~$k$- and $\Omega$-dependence of the FL exponent obtained by numerically solving Eq.~(\ref{eq:S}) with $(q_0, q_\mathrm{osc}, c_1\bar{\rho})=(0.5, 0.39, -1/20)c_0\bar{\rho}$. The dotted curves depict the analytically obtained resonance condition $\hbar\Omega=2E_k/J$ for $J=1,2,3$, and 4. Although the resonance at $J=4$ is not so visible in this color scale, $\lambda_{\rm FL}$ is nonzero in a vicinity of $\hbar\Omega=2E_k/4$. Note that plotting $\lambda_\mathrm{FL}$ at $\Omega\ll E_k/\hbar$ is difficult because the intervals between the resonant lines $2E_k[1/J-1/(J+1)]$ become infinitesimally small. }
			\label{FIG:intsec}
		\end{center}
	\end{figure}

	\subsection{Resonant conditions}
	\label{sec:resonant_conditions}
	We start from the simplest case of $b=0$, at which the matrix~\eqref{matrix:Q} becomes diagonal and has the eigenvalues
	\begin{align}
	i\lambda_\mathrm{F}=\omega_{j,\pm}\equiv j\Omega\pm \frac{E_k}{\hbar} \ (j\in\mathbb{Z}).
	\label{eq:lambda_F_b=0}
	\end{align}
    Note that all $\lambda_\mathrm{F}$'s are pure imaginary in this case, which means that the amplitudes of the spin density vectors and nematic density tensors do not grow in time.
	
	A small finite $b$ couples the above eigenmodes, and its effect becomes prominent when two eigenvalues are close to each other. Note that $\hat{Q}$ in Eq.~\eqref{eq:EigenEq} is a pseudo-Hermitian matrix, i.e., there exists an Hermitian matrix $\eta$ such that $ \hat{Q}^\dagger=\eta^\dagger\hat{Q}\eta$~\cite{Kato.ep}; In the present case, $\eta$ is given by $\eta=\textrm{Diag}\,[\cdots,1,-1,1,-1,\cdots]$. In such a case, two close eigenvalues can coalesce and become a pair of complex conjugate values.
	Thus, $\lambda_\mathrm{F}$'s have nonzero real parts around the intersections of $\omega_{j,\pm}~(j=0,\pm1,\pm2,\cdots)$  as a function of $\Omega$ [see Fig.~\ref{FIG:intsec}(a)]. By solving $\omega_{j_1,+}=\omega_{j_2,-}$ with $j_1,j_2\in\mathbb{Z}$, we obtain the intersection points at
	\begin{align}
		\label{intsec}
	\Omega=\displaystyle{\frac{2E_k}{J\hbar}},
	\end{align}
	where $J\equiv j_2-j_1$. Since $\Omega$ is assumed to be positive, we restrict $J$ to be a positive integer. Equation~\eqref{intsec} is the approximated condition for the Shapiro resonance with spatial degrees of freedom.
	
	To confirm the above argument, we numerically calculate $\lambda_\mathrm{FL}$ from Eq.~$(\ref{eq:S})$ by following the method used in Ref.~\cite{lambdaNumCal}: For a fixed wavenumber, we numerically solve Eq.~\eqref{eq:S} from $t=0$ to $T$ with the initial conditions $(S_A,S_B)^\mathrm{T}=(1,0)^\mathrm{T}$ and $(0,1)^\mathrm{T}$, obtaining $\bm{S}_1(T)$ and $\bm{S}_2(T)$, respectively; The $2\times 2$ matrix $\hat{U}=(\bm{S}_1(T), \bm{S}_2(T))$ is the one-period time-evolution operator and its eigenvalue $\lambda_U$ is related to the Floquet exponent as $\lambda_U=e^{\lambda_\textrm{F}T}$; Thus, we obtain the FL exponent from the numerically obtained $\lambda_U$ as $\lambda_\textrm{FL}=\textrm{Re}\,[(\ln \lambda_U)/T]$.
	The color plot in Fig.~\ref{FIG:intsec}(b) shows the numerical result of $\lambda_\mathrm{FL}$, where we obtain two FL exponents with opposite signs and plot only the positive one in Fig.~\ref{FIG:intsec}(b). We also plot Eq.~\eqref{intsec} with dotted curves in the same figure, which show an excellent agreement with the numerical result.
	
	Note that although Eq.~\eqref{eq:EigenEq} has an infinite number of eigenvalues, we obtain only a pair of FL exponents in the numerical procedure, suggesting that the imaginary part of the eigenvalues of $\hat{Q}$ can only take two values corresponding to the numerically obtained $\lambda_\mathrm{FL}$. Indeed, we will see below (and also in Appendix~\ref{app3_Q}) that the detailed analysis of the infinite-dimensional matrix $\hat{Q}$ results in a pair of FL exponents with opposite signs.
	
	\begin{figure}[t]
		\begin{tabular}{c}
			\begin{minipage}[t]{1.0\hsize}
				\includegraphics[keepaspectratio, width=8.5cm,clip]{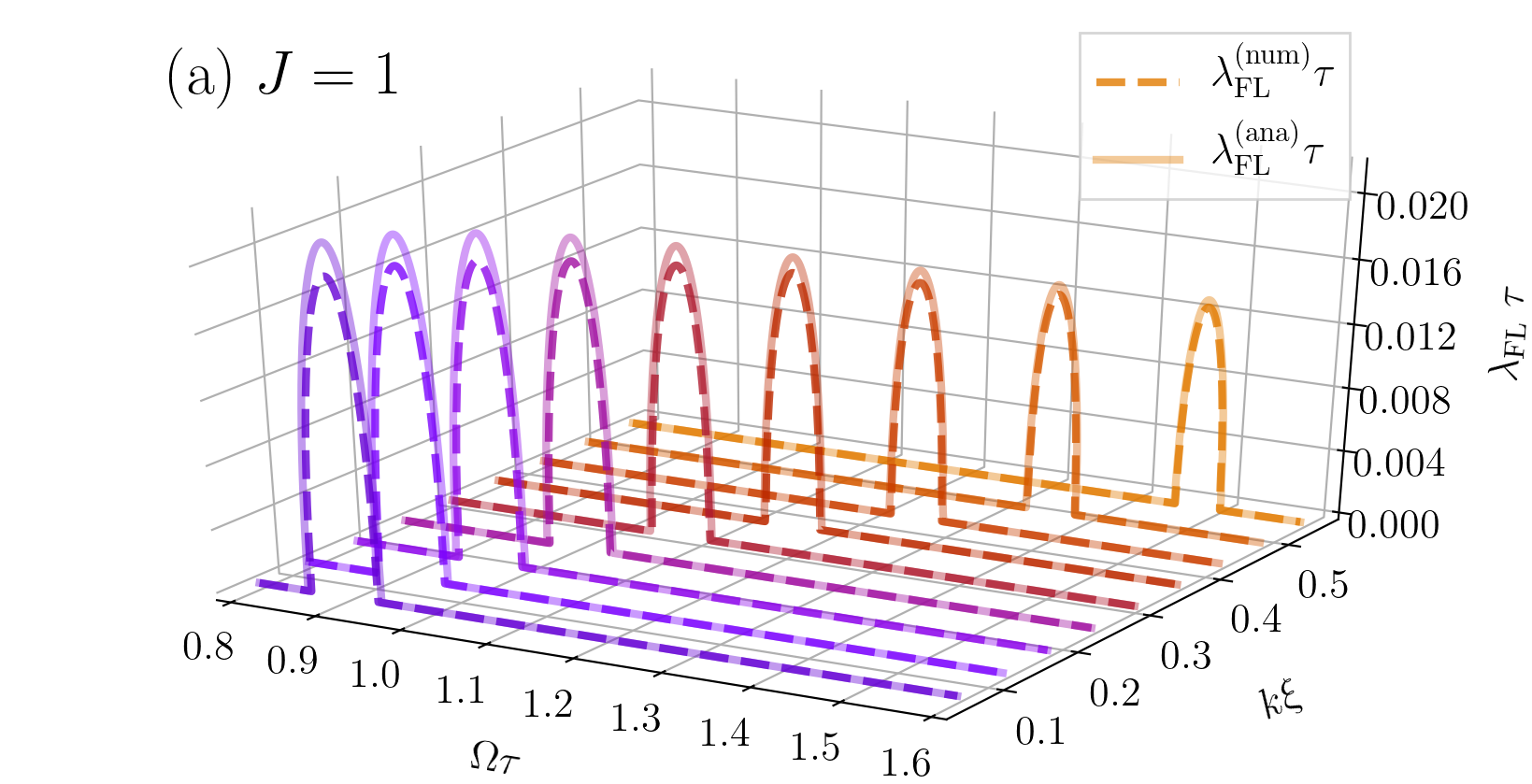}
				\includegraphics[keepaspectratio, width=8.5cm,clip]{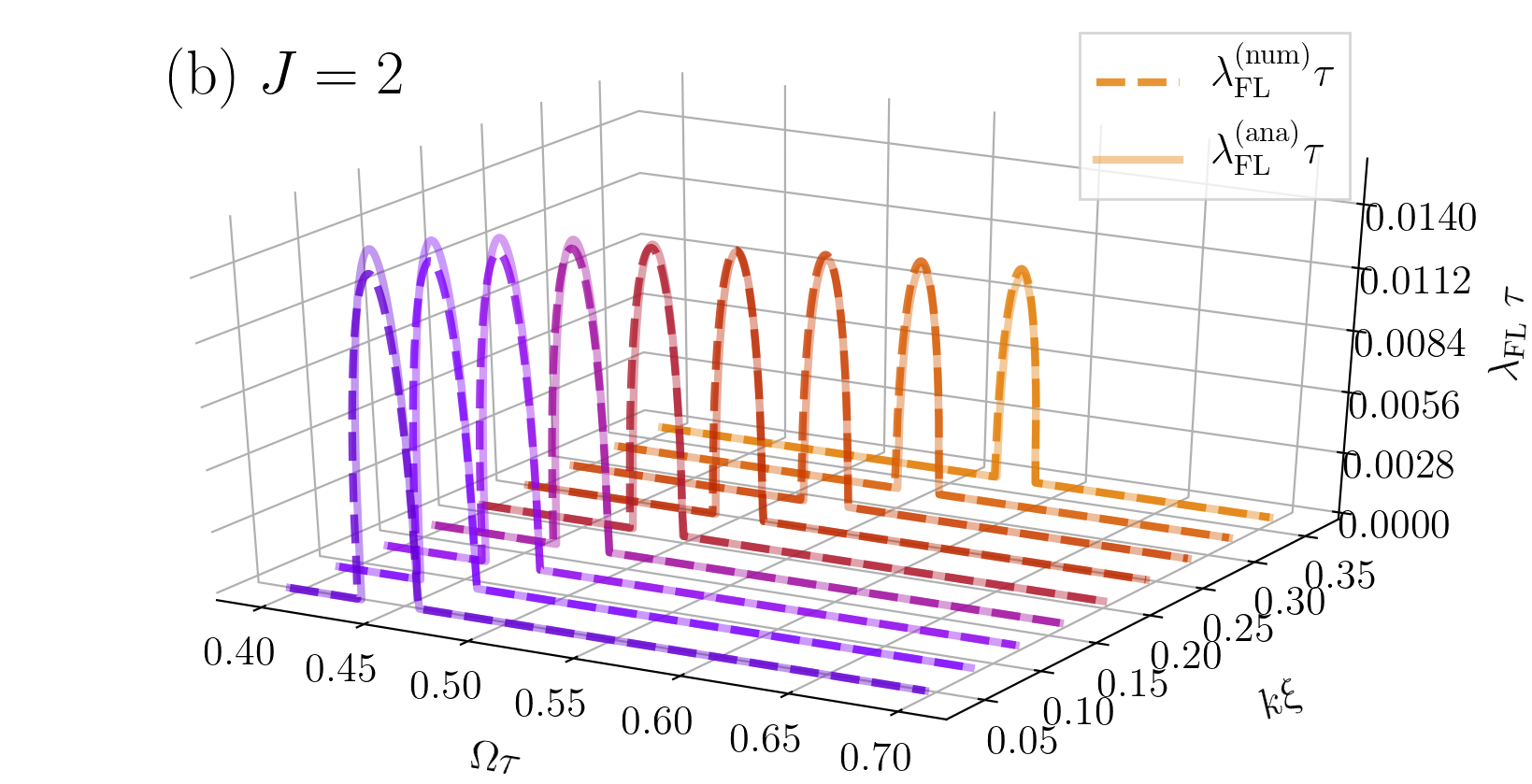}
			\end{minipage}
		\end{tabular}
		\caption{Comparison between $\lambda_\mathrm{FL}^\mathrm{(ana)}$ and $\lambda_\mathrm{FL}^\mathrm{(num)}$ for (a) $J=1$ and (b) $J=2$. Here, $\lambda_\mathrm{FL}^\mathrm{(ana)}$ is the analytical result given by Eqs.~\eqref{lambFL1} and \eqref{eq:FDMAJ=2} for (a) and (b), respectively, and $\lambda_\mathrm{FL}^\mathrm{(num)}$ is the numerically obtained one shown in Fig.~\ref{FIG:intsec}(b). The parameters are the same as those in Fig.~\ref{FIG:intsec}(b). For a better visibility, we color the curves according to the value of $k\xi$. We can see that the FDMA works well, in particular, at the large wavenumbers.}\label{FIG:lamNum}
	\end{figure}

	\subsection{Finite-dimensional matrix approximation \label{subsec:FDMA}} 
	We have obtained the resonant condition of Eq.~\eqref{intsec}, but our numerical results in Fig.~\ref{FIG:intsec}(b) show that there is a width of the resonant frequency. In this section, we analytically calculate the FL exponent and the width of the resonance by approximating the infinite-dimensional matrix in Eq.~\eqref{eq:EigenEq} with a finite one \cite{Salwen1955,Shirley1965}. 
	
	Our procedure is as follows. Let $\ket{C_{j,\pm}}$ denote the normalized eigenmode at $b=0$ with eigenvalue $\omega_{j,\pm}$. We consider the case when the oscillating frequency is close to the $J$th resonance, $\hbar\Omega\simeq 2E_k/J$, at which $\ket{C_{j,+}}$ and $\ket{C_{j+J,-}}$ are almost degenerate. In such a case, we can neglect the other modes, and $\hat{Q}$ is approximated by the $2\times 2$ matrix $\bra{a}\hat{Q}\ket{a'}$ with $a,a'=C_{j,+}$ and $C_{j+J,-}$.
	This procedure is similar to what we do in calculating the energy bands of the nearly free electron model, where a band gap opens at the boundary of the Brillouin zone.
	However, the $2\times 2$ matrix is not enough for $J>1$ because the off-diagonal elements  vanish for $J>1$; 
	In the perturbative expansion of $\hat{Q}$ in powers of $b$, the coupling between $\ket{C_{j,+}}$ and $\ket{C_{j+J,-}}$ first appears in the $J$th order term. 
	We therefore need to take into account the intermediate states that appear in the coupling between $\ket{C_{j,+}}$ and $\ket{C_{j+J,-}}$ and approximate $\hat{Q}$ with the one projected onto the restricted Hilbert space. Below, we demonstrate the cases of $J=1$ and $2$.

	\subsubsection{$J=1$}
	We consider the coupling between $\ket{C_{j,+}}$ and $\ket{C_{j+1,-}}$ which is the first order in $b$. We approximate the matrix $\hat{Q}$ with the $2\times 2$ matrix:
	\begin{align}
		\label{eq:FDMAJ=1}
		&\begin{pmatrix}
		\bra{C_{j,+}} \hat{Q}\ket{C_{j,+}} & \bra{C_{j,+}} \hat{Q}\ket{C_{j+1,-}}\\
		\bra{C_{j+1,-}} \hat{Q}\ket{C_{j,+}} & \bra{C_{j+1,-}} \hat{Q}\ket{C_{j+1,-}}
		\end{pmatrix} \nonumber\\
		&=\begin{pmatrix} (\hat{F}_j)_{2,2} & -b(\hat{G})_{2,1} \\ b(\hat{G})_{1,2} &(\hat{F}_{j+1})_{1,1} \end{pmatrix} \nonumber\\
		&=\frac{1}{\hbar} \begin{pmatrix} j\hbar\Omega+E_k & -ibc_1\bar{\rho} \\ -ibc_1\bar{\rho}
		&(j+1)\hbar\Omega-E_k \end{pmatrix},
	\end{align}
	from which we obtain the FL exponent
	\begin{align}
		\label{lambFL1}
		\lambda_\mathrm{FL}^{\textrm{(ana)}}=\mathrm{Re}\left[\ \frac{1}{2\hbar}\sqrt{\left(\frac{q_\mathrm{osc}c_1\bar{\rho}}{E_k}\right)^2-\left(\hbar\Omega-2E_k\right)^2}\ \right].
    \end{align}
	The FL exponent takes a nonzero value when the frequency $\Omega$ is in
	the resonant region:
	\begin{align}
		\label{ieq:1}-\frac{q_\mathrm{osc}|c_1|\bar{\rho}}{E_k}<\hbar\Omega-2E_k<\frac{q_\mathrm{osc}|c_1|\bar{\rho}}{E_k}.
	\end{align}
	Note that although $j$ takes an arbitrary integer, the $j$ dependence of the eigenvalue appears only in the real part of $i\lambda_\textrm{F}$. 
	That is, all the intersection points in Fig.~\ref{FIG:intsec}(a) at $\hbar\Omega=2E_k$ give different $\textrm{Im}[\lambda_\textrm{F}]$'s but the same FL exponent.
	This is consistent with the fact that we have only two FL exponent in the numerical calculation.
	
	To see the validity of the analytical result~\eqref{lambFL1}, we compare $\lambda_\mathrm{FL}^{(\mathrm{ana})}$ and $\lambda_\mathrm{FL}^{(\mathrm{num})}$ in Fig.~\ref{FIG:lamNum}(a), where $\lambda_\mathrm{FL}^\mathrm{(num)}$ is the numerically obtained FL exponent shown in Fig.~\ref{FIG:intsec}(b).
	One can see that the analytical result well reproduces the width and the amplitude of the resonance obtained by the numerical calculation.
	
	\subsubsection{$J=2$}
	In the case of $J=2$, we consider the coupling between $\ket{C_{j-1,+}}$ and $\ket{C_{j+1,-}}$. These states couple with each other via the intermediate state $\ket{C_{j,\pm}}$ in the second order of $b$. Note that in the same order of $b$, the couplings with $\ket{C_{j-2,\pm}}$ and $\ket{C_{j+2,\pm}}$ states shift the eigenvalues associated with $\ket{C_{j-1,+}}$ and $\ket{C_{j+1,-}}$, respectively. Thus, we need to solve the following $8\times 8$ matrix eigenvalue problem:
	\begin{widetext}
		\begin{align}
		\left|\left(\begin{array}{cccccccc}
		(\hat{F}_{j-2})_{1,1} & 0 & -b(\hat{G})_{1,2} & 0 & 0 & 0 & 0 & 0\\
		0 & (\hat{F}_{j-2})_{2,2} & -b(\hat{G})_{2,2} & 0 & 0 & 0 & 0 & 0\\
		b(\hat{G})_{2,1} & b(\hat{G})_{2,2} & (\hat{F}_{j-1})_{2,2} & -b(\hat{G})_{2,1} & -b(\hat{G})_{2,2} & 0 & 0 & 0\\
		0 & 0 & b(\hat{G})_{1,2} & (\hat{F}_j)_{1,1} & 0 & -b(\hat{G})_{1,1} & 0 & 0\\
		0 & 0 & b(\hat{G})_{2,2} & 0 & (\hat{F}_j)_{2,2} & -b(\hat{G})_{2,1} & 0 & 0\\
		0 & 0 & 0 & b(\hat{G})_{1,1} & b(\hat{G})_{1,2} & (\hat{F}_{j+1})_{1,1} & -b(\hat{G})_{1,1} & -b(\hat{G})_{1,2}\\
		0 & 0 & 0 & 0 & 0 & b(\hat{G})_{1,1} & (\hat{F}_{j+2})_{1,1} & 0\\
		0 & 0 & 0 & 0 & 0 & b(\hat{G})_{2,1} & 0 & (\hat{F}_{j+2})_{2,2}
		\end{array}\right)-i\lambda_\mathrm{F}\hat{I}\ \right|=0.
    \label{eq:EigenEq_J=2}
		\end{align}
	\end{widetext}
	The analytical expression of the eigenvalue is so complicated that we show the detail in Appendix \ref{app5_J=2}. We compare the obtained analytical solution $\lambda_\textrm{FL}^\textrm{(ana)}$ with the numerical one in Fig.~\ref{FIG:lamNum}(b). As in the case of $J=1$, one can see that $\lambda_\textrm{FL}^\textrm{(ana)}$ exhibits excellent agreement with the numerical result.

	\subsection{FL exponent on resonance}

	\begin{figure}[t]
		\begin{center}
			\includegraphics[keepaspectratio, width=8.5cm,clip]{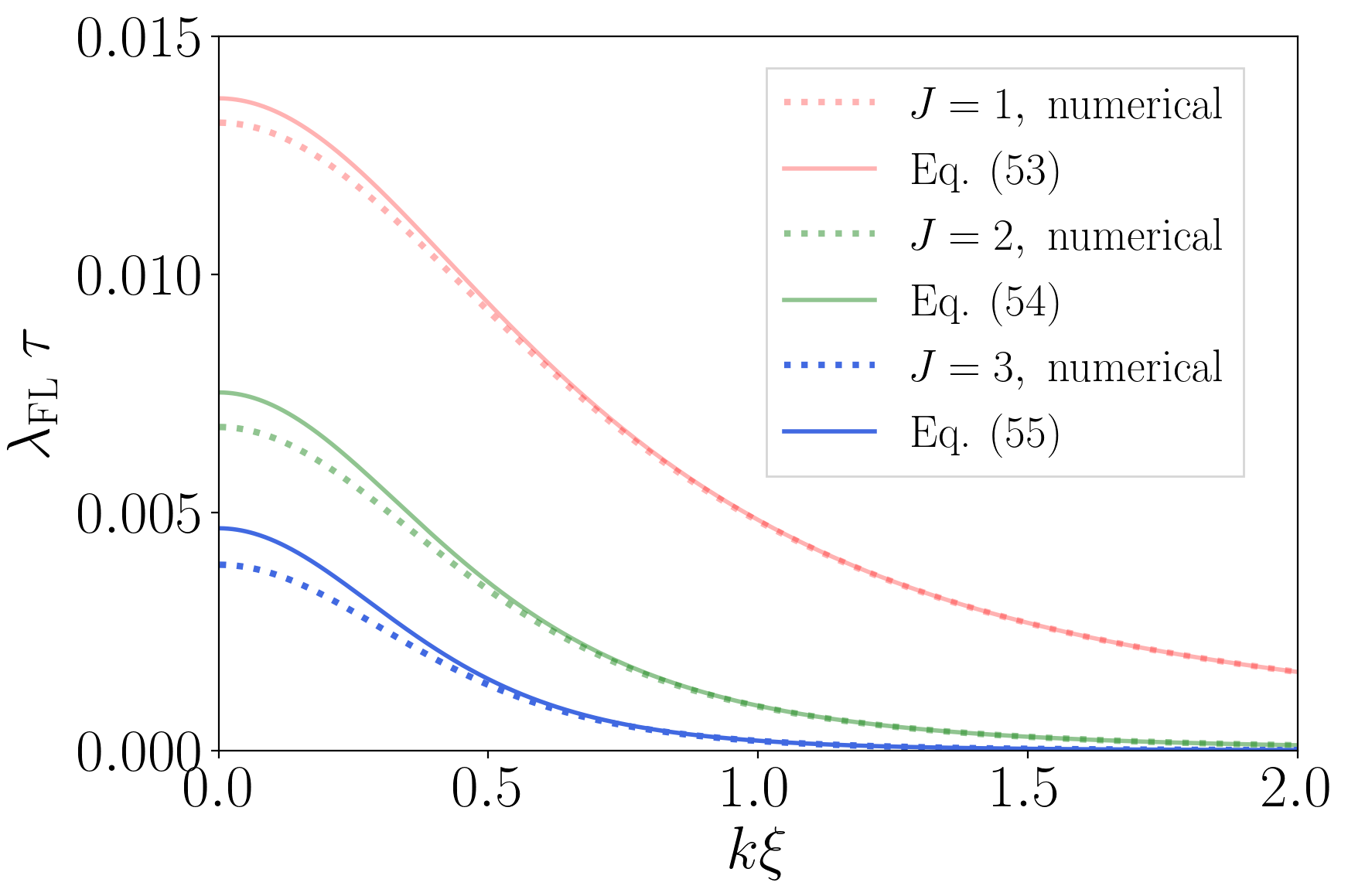}
			\caption{Floquet-Lyapunov exponents at $\hbar\Omega=2E_k/J$ with $J=1, 2$, and $3$ as functions of $k\xi$, where the dotted curves show the numerical results and the solid curves depict Eqs.~\eqref{eq:lam1}-\eqref{eq:lam3}.
			The parameters are $(q_0, q_\mathrm{osc},c_1\bar{\rho})=(0.5, 0.3,+1/20)c_0\bar{\rho}$ . The analytical results well reproduce the numerical ones, in particular, at the large wavenumbers.}
			\label{FIG:ptb}
		\end{center}
	\end{figure}

    Although we can calculate $\lambda_\mathrm{FL}$ for higher $J$ in a similar manner as in the cases of $J=1$ and 2, it requires tedious calculations. We note that $\lambda_\mathrm{FL}$ at $\hbar\Omega=2E_k/J$ can be obtained more easily by using the conventional perturbation theory for degenerate states. We regard $b$ as a small perturbation parameter and rewrite Eq.~$(\ref{eq:EigenEq})$ in the following form:
	\begin{align}
		\label{eq:ptb}\left(\hat{E}+b\hat{V}\right)\ket{C}=i\lambda_\mathrm{F}\ket{C},
	\end{align}
	where $\hat{E}$ and $b\hat{V}$ are the matrices composed of the diagonal and off-diagonal elements of $\hat{Q}$, respectively. As we have discussed in the above, at $\hbar\Omega=2E_k/J$, the states $\ket{C_{j,+}}$ and $\ket{C_{j+J,-}}$ are degenerate at $b=0$. These states are coupled in the $J$th order perturbation, giving rise to a nonzero $\lambda_\mathrm{FL}$. Following the conventional perturbation theory for degenerate states, $\lambda_\mathrm{FL}$'s for $J=1, 2$ and $3$ are obtained as the imaginary parts of the eigenvalues of the $2\times 2$ matrices, respectively, given by
	\begin{align}
	    V_{ll'}^{(J=1)}&=b\bra{ C_l}\hat{V}\ket{C_{l'}},\\
	    V_{ll'}^{(J=2)}&=b^2 \sum_{l_1}
	    \frac{\bra{ C_l}\hat{V}\ket{C_{l_1}}\bra{C_{l_1}}\hat{V}\ket{C_{l'}}}{\omega_{l}-\omega_{l_1}},
	    \label{eq:Vll^J=2}\\
	    V_{ll'}^{(J=3)}&=b^3 \sum_{l_1,l_2}
	    \frac{\bra{ C_l}\hat{V}\ket{C_{l_1}}\bra{C_{l_1}}\hat{V}\ket{C_{l_2}}\bra{C_{l_2}}\hat{V}\ket{C_{l'}}}{(\omega_{l}-\omega_{l_1})(\omega_{l}-\omega_{l_2})},
	    \label{eq:Vll^J=3}
	\end{align}
    where $l,l'=(j,+)$ and $(j+J,-)$, and we take the summation over $l_1, l_2=(j,s)$ for all possible combinations of $j\in\mathbb{Z}$ and $s=+$ and $-$ other than $(j,+)$ and $(j+J,-)$. The resulting FL exponents are given by
    \begin{align}
		\label{eq:lam1}\lambda^{(J=1)}_\mathrm{FL}&=\frac{q_\mathrm{osc}|c_1|\bar{\rho}}{2\hbar E_k},
		\\\label{eq:lam2}\lambda^{(J=2)}_\mathrm{FL}&=\frac{q_\mathrm{osc}^2|c_1|\bar{\rho}}{6\hbar E_k^3}\sqrt{9E_k^2+5\left(c_1\bar{\rho}\right)^2},
		\\\label{eq:lam3}\lambda^{(J=3)}_\mathrm{FL}&=\frac{9q_\mathrm{osc}^3|c_1|\bar{\rho}}{128\hbar E_k^5}\left(8E_k^2+9\left( c_1\bar{\rho}\right)^2\right).
	\end{align}
    The results for $J=1$ and 2 coincide with Eqs.~\eqref{lambFL1} and \eqref{eq:FDMAJ=2}, respectively. We can calculate the FL exponent for $J\ge 4$ in the same manner, but we need to be careful that the diagonal terms of $V_{ll'}^{(J')}$ with even $J' < J$ generally lifts the degeneracy of $(j,+)$ and $(j+J,-)$ states, which also contributes to $\mathrm{Im}\lambda_\mathrm{F}$ for $J=3$.

	Figure~\ref{FIG:ptb} compares numerical results of $\lambda_\mathrm{FL}$ (dotted lines) with Eqs.~(\ref{eq:lam1})--(\ref{eq:lam3}) (solid lines). As one can see, the approximation works quite well for larger $k\xi$ because the perturbation parameter $b=q_\mathrm{osc}/2E_k$ becomes smaller for larger $k\xi$.

	\section{Nonlinear resonant dynamics \label{Sec:num}}
	We numerically solve the spin-1 spinor GPE~(\ref{eq:GP3D}) in a one-dimensioinal system and investigate dynamics of the Shapiro resonance beyond the linear analysis. We first confirm the validity of the linear analysis in Sec.~\ref{sec:linear}. We then study the nonlinear dynamics using the hydrodynamic variables and discuss how the long-time dynamics proceeds.
	
	\begin{figure}[t]
		\begin{center}
			\includegraphics[keepaspectratio, width=8.5cm,clip]{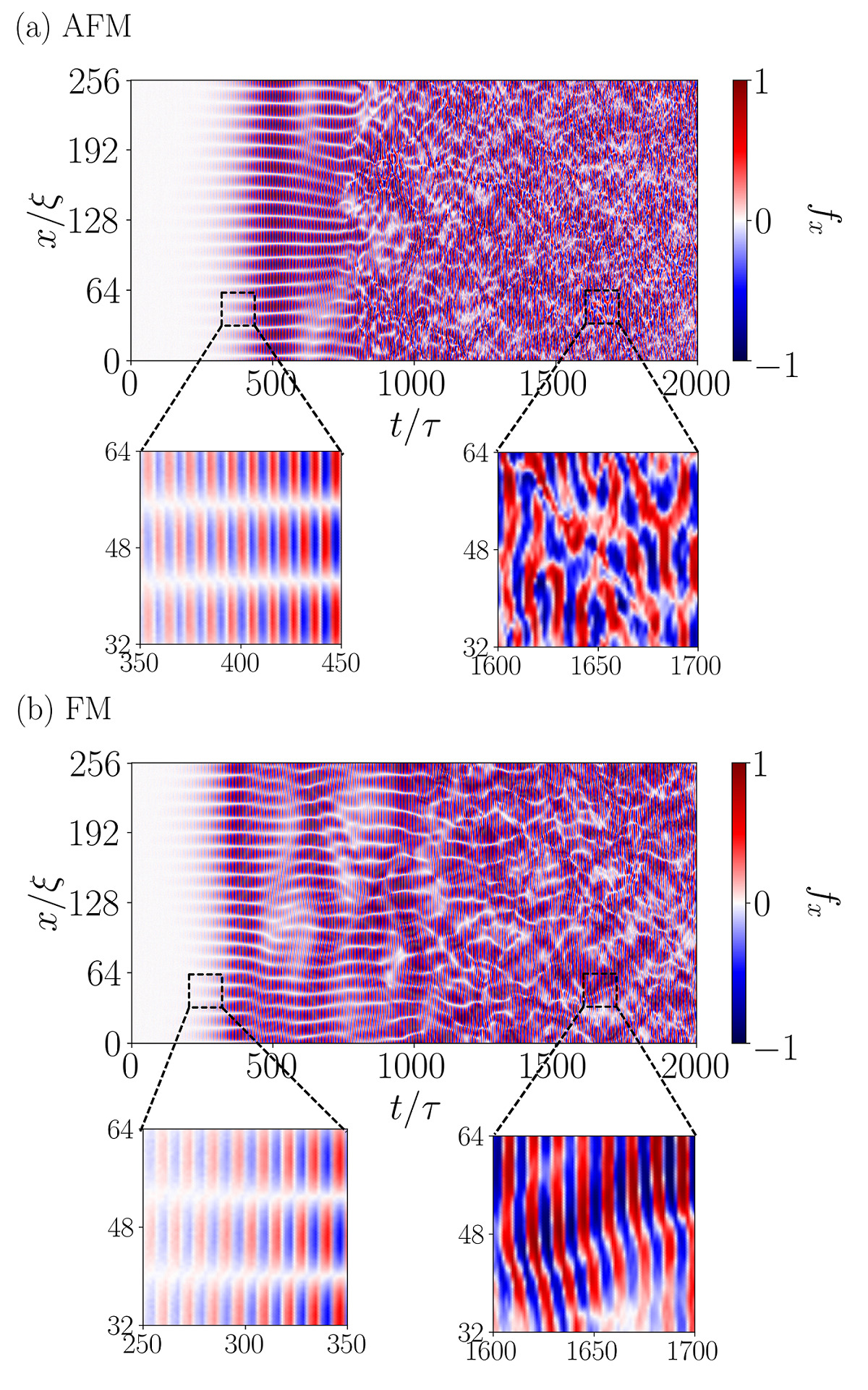}
			\caption{
				Time evolution of spin density $f_x$ in (a) antiferromagnetic (AFM) and (b) ferromagnetic (FM) systems, for which we use the spin-dependent interaction $c_1/c_0=1/20$ and $-1/20$, respectively. The parameters for the QZ terms are $(q_0, q_{\rm osc}, \hbar \Omega)=(0.5c_0\bar{\rho}, 0.39c_0\bar{\rho}, 2E_{k_{\rm res}})$. The insets show the enlarged view of the early and late stages.	 The spin density in the AFM system tends to be finer structure than that in the FM one. The growing speed of $f_x$ in the AFM system is slightly slower than that in the FM one because $\lambda_{\rm FL }$ of the former is smaller than that of the latter. 
			}
			\label{FIG:GPEtimeEvo}
		\end{center}
	\end{figure}

	\subsection{Numerical results using the GPE}
	We solve Eq.~(\ref{eq:GP3D}) for a 1D system of a system size $L_x/\xi=256$ with a periodic boundary condition. The initial wavefunction is given by
\begin{align}
	\label{eq:initial}\psi_m(x,t=0)=\sqrt{\bar{\rho}}\delta_{m,0}+\frac{\sqrt{\bar{\rho}}}{100}R_{m,1}(x)e^{i2\pi R_{m,2}(x)},
\end{align}
    where $R_{m,l}(x)$ $(l=1,2)$ is a uniform random number in [0,1]. We choose the parameters as $(q_0,q_\mathrm{osc}, \hbar\Omega)=(0.5, 0.39, 2E_{k_\mathrm{res}})$ with $k_\mathrm{res}=10\times2\pi/L_x$. According to Eq.~(\ref{lambFL1}), the $\Omega$ is the resonance frequency of $J=1$, and the spin waves with the wavelength $L_x/10$ will grow. The interaction parameters are set to be $c_1/c_0=1/20$ and $-1/20$ for AFM and FM BECs, respectively.

	In Fig.~\ref{FIG:GPEtimeEvo}, we plot the time evolution of the spin density $f_x(x,t)$ which is expected to grow as $e^{\lambda_\textrm{FL}t}$ within the linear analysis. Figures~\ref{FIG:GPEtimeEvo}(a) and (b) show the spatial and temporal distributions of $f_x(x,t)$ for the AFM and FM systems, respectively. As expected from the linear analysis, the spin waves with wavelength $L_x/10$ grow in the early stage (see the enlarged views in Fig.~\ref{FIG:GPEtimeEvo}). In the late stage ($t/\tau\gtrsim 1500$), $f_x(x,t)$ in the AFM system exhibits finer structure than that in the FM system. This can be seen more clearly in the $k$-space.

\begin{figure}[t]
		\begin{tabular}{c}
			\begin{minipage}[t]{1.0\hsize}
				\includegraphics[keepaspectratio, width=8.5cm,clip]{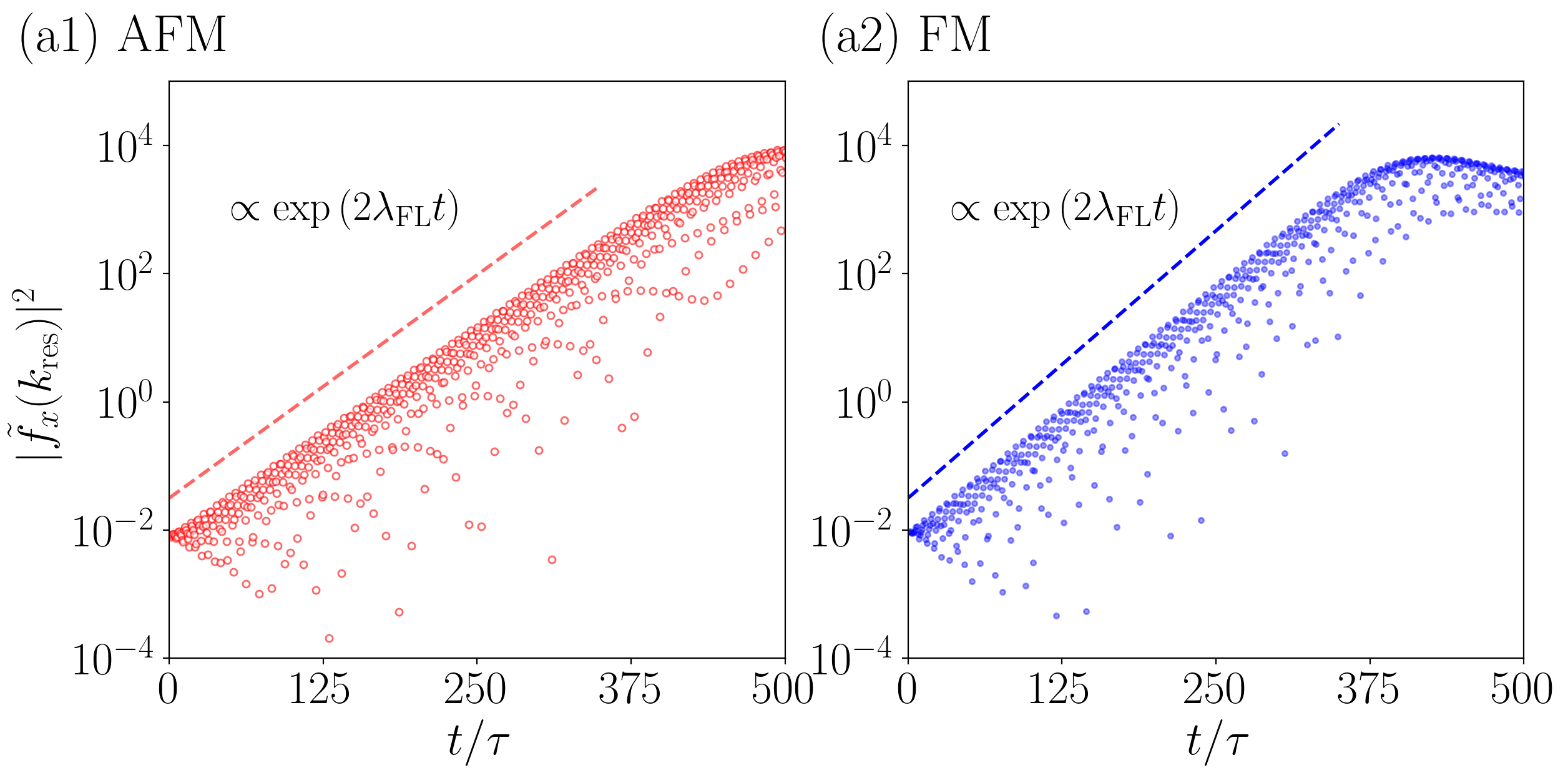}
			\end{minipage}\\
			\begin{minipage}[t]{1.0\hsize}
				\includegraphics[keepaspectratio, width=8.0cm,clip]{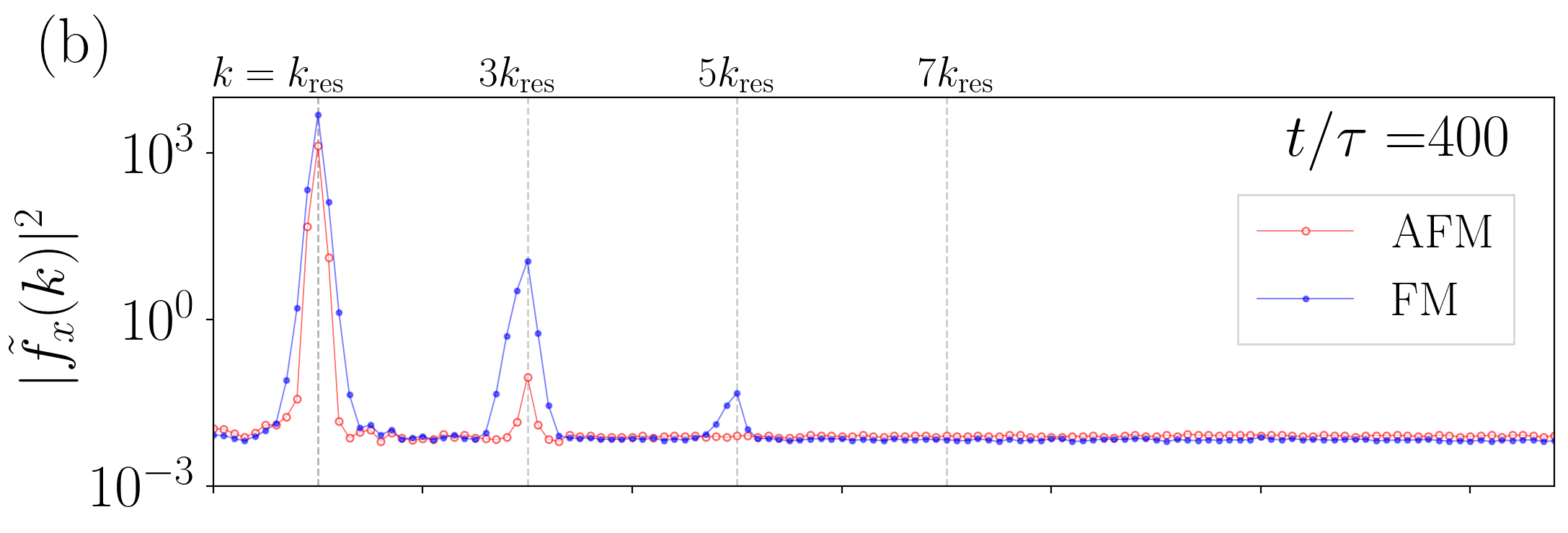}
			\end{minipage}\\
			\begin{minipage}[t]{1.0\hsize}
				\includegraphics[keepaspectratio, width=8.0cm,clip]{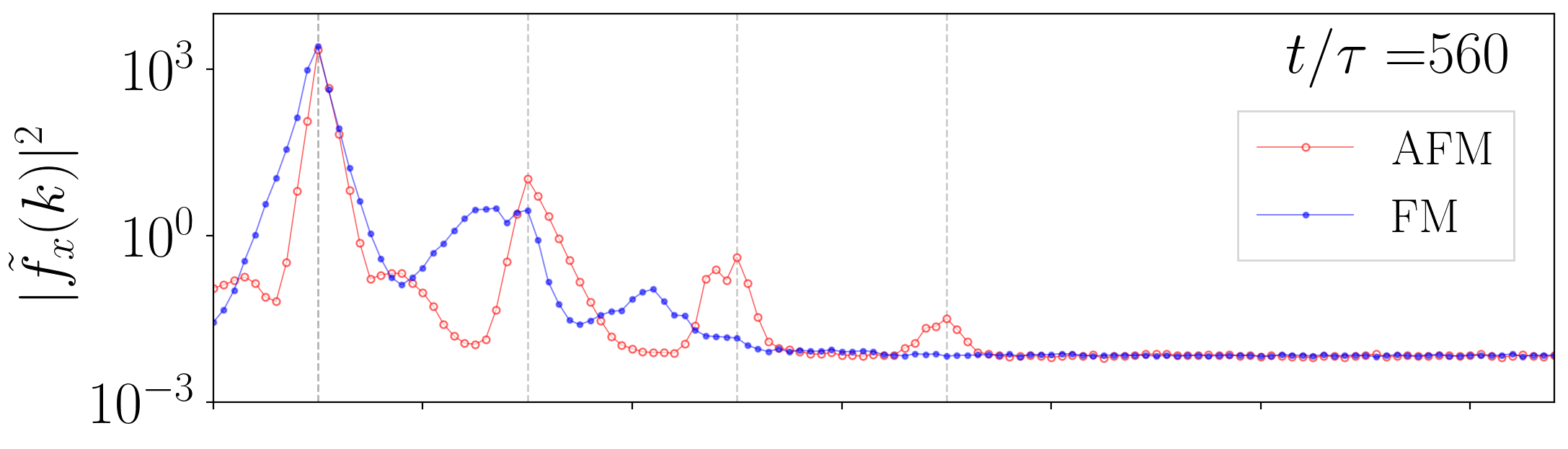}
			\end{minipage}\\
			\begin{minipage}[t]{1.0\hsize}
				\includegraphics[keepaspectratio, width=8.0cm,clip]{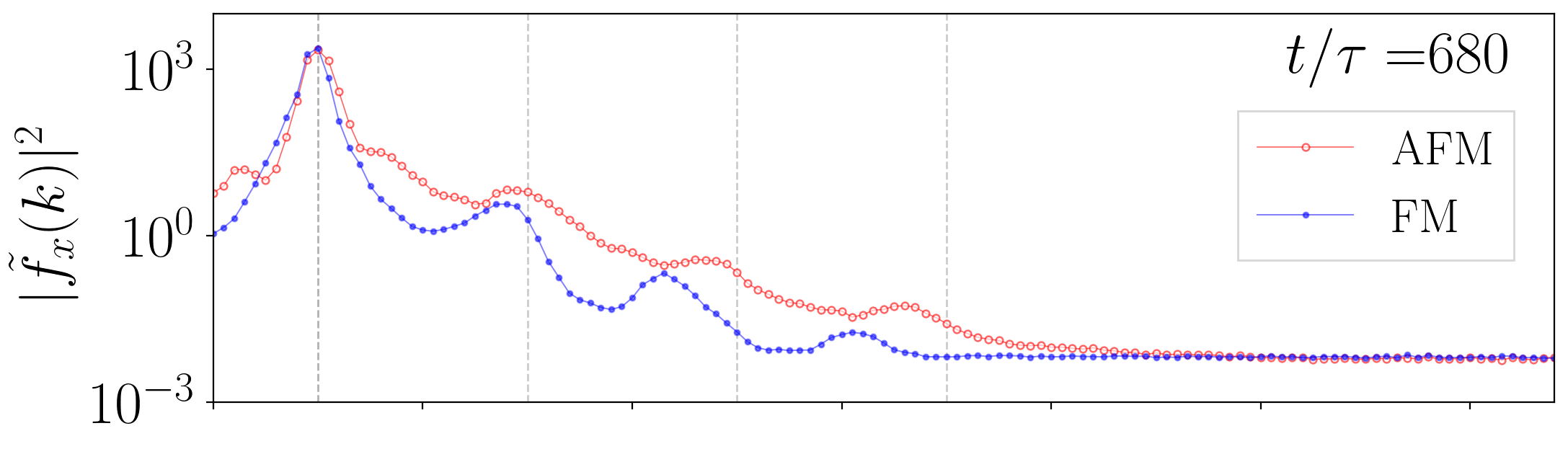}
			\end{minipage}\\
			\begin{minipage}[t]{1.0\hsize}
				\includegraphics[keepaspectratio, width=8.0cm,clip]{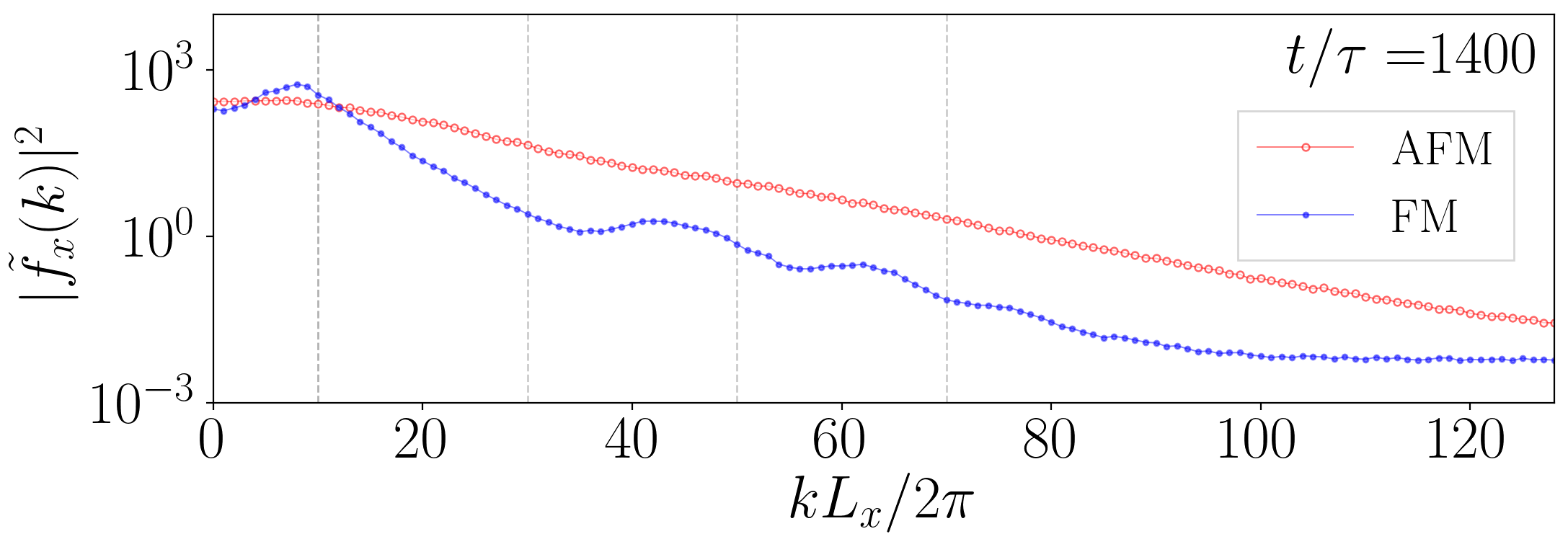}
			\end{minipage}\\
		\end{tabular}
		\caption{(a1),(a2) Time evolution of $|\tilde{f}_x(k_\mathrm{res},t)|^2$ in the AFM and FM systems. The parameters used here are the same as those of Fig.~\ref{FIG:GPEtimeEvo}. According to our linear analysis, $|\tilde{f}_x(k_\mathrm{res},t)|^2$ grows with $\exp(2\lambda_\mathrm{FL}t)$, and one can see that the dashed lines proportional to $\exp(2\lambda_\mathrm{FL}t)$ are parallel to the numerical data. 
        (b) $k$-space spectra of the spin density $f_x(x,t)$ in $t/\tau=400, 560, 680,$ and $1400$. The parameters are the same as those of (a1) and (a2). The vertical dashed lines are at $k=k_\mathrm{res}, 3k_\mathrm{res}, 5k_\mathrm{res}$, and $7k_\mathrm{res}$. All data are obtained by averaging over 1000 samples with different initial noises.}\label{FIG:k-space}
	\end{figure}

	Figure~\ref{FIG:k-space} shows the dynamics in the $k$-space. Here, we define the Fourier components in the 1D system as $\tilde{A}(k)\equiv\int_0^{L_x} A(x)e^{-ikx}dx$ and plot $|\tilde{f}_x(k,t)|^2$ averaged over 1000 samples of random initial states given by Eq.~\eqref{eq:initial}.  Figures~\ref{FIG:k-space}(a1) and (a2) show the time evolution of $|\tilde{f}_x(k_\mathrm{res},t)|^2$ for the AFM and FM systems, respectively, where the dashed lines are the analytical result $e^{2\lambda_\textrm{FL}t}$ with  $\lambda_\textrm{FL}$ given in Eq.~\eqref{eq:lam1}. One can clearly see that, for both cases, $|\tilde{f}_x(k_\textrm{res},t)|^2$ in the early stage agrees well with the linear analysis. 
	
	The linear analysis breaks down in the long-time dynamics, and $|\tilde{f}_x(k_\textrm{res},t)|^2$ in Figs.~\ref{FIG:k-space} (a1) and (a2) deviates from the exponential growth. The breakdown of the linear analysis is also seen in the $k$ dependence of $|\tilde{f}_x(k,t)|^2$ shown in Fig.~\ref{FIG:k-space}(b), where the results for AFM (FM) systems are shown with open (filled) circles. As one can see, in addition to the main peak at $k=k_\textrm{res}$, small peaks appear around odd multiples of $k_\mathrm{res}$ at $t/\tau=400$. They are consequences of the nonlinear effect as mentioned in the next subsection. As time evolves, these peaks become broader, and $|\tilde{f}_x(k,t)|^2$ eventually distributes in the wide range of $k$. Figure~\ref{FIG:k-space}(b) directly shows that the sign of $c_1$ significantly alters the late dynamics especially at $t/\tau\ge680$: The spectra of the AFM system have the larger values in the high wavenumber region than those of the FM system. This is consistent with the real space configuration of $f_x(x,t)$ in Fig.~\ref{FIG:GPEtimeEvo}.
	
	\begin{figure*}[t]
		\begin{center}
			\includegraphics[keepaspectratio, width=17.5cm,clip]{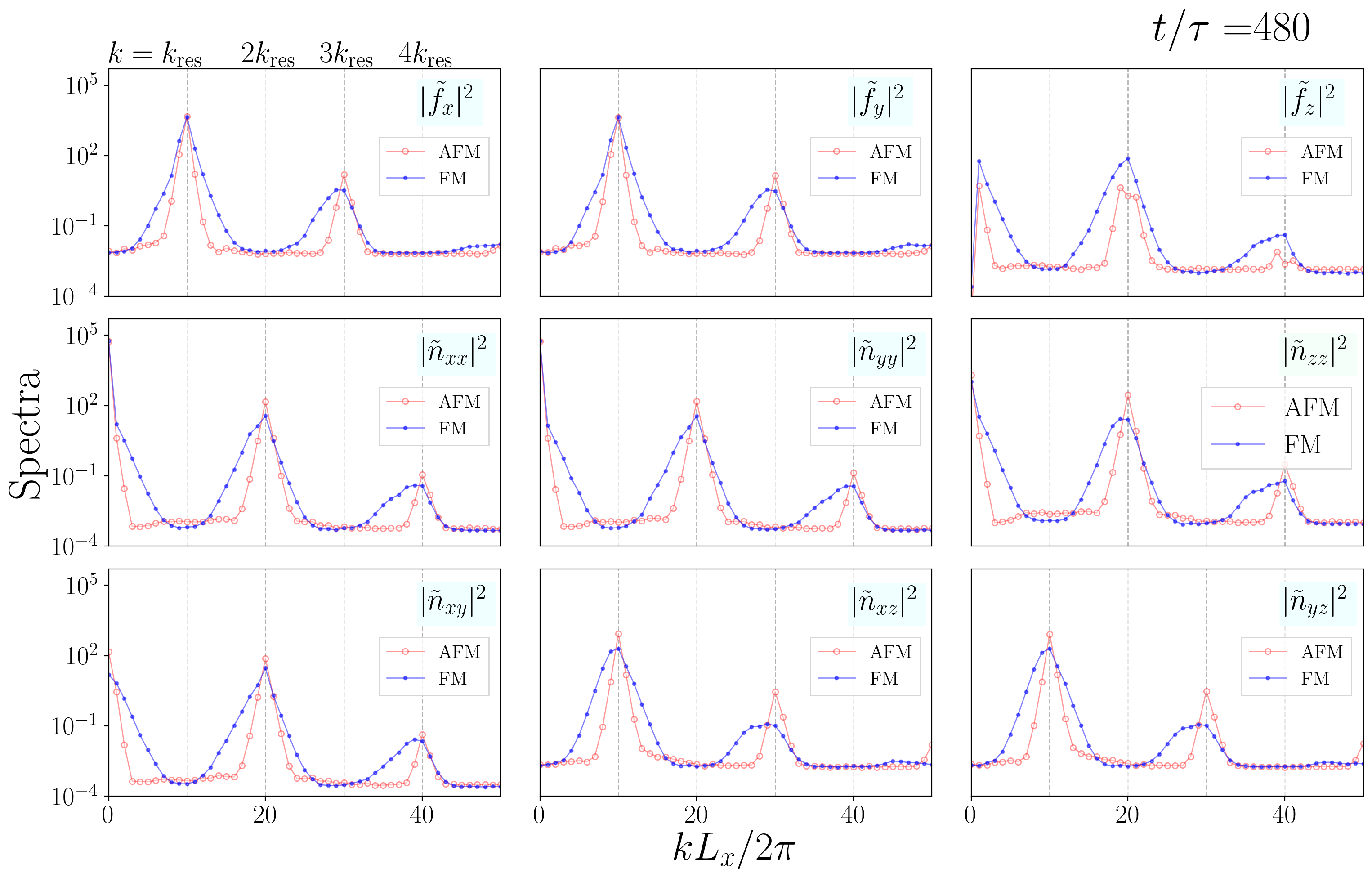}
			\caption{$k$-space spectra of all the spin density vectors and nematic density tensors at $t/\tau=480$. The parameters used here is the same as those of Fig.~\ref{FIG:k-space}. The open and filled circles are for the AFM and FM systems, respectively. As described in the main text, the resonant variables $(\tilde{f}_x, \tilde{f}_y, \tilde{n}_{xz}$, and $\tilde{n}_{yz})$ have peaks around odd multiples of $k_\mathrm{res}$ while others have peaks around even multiples of $k_\mathrm{res}$. The magnitude of peaks in $\delta f_z$ and $\delta n_{zz}$ are consistent with the order we estimate from Eqs.~(\ref{eq:kdel_fz}) and (\ref{eq:kdel_nzz}), respectively.}
			\label{FIG:k9}
		\end{center}
	\end{figure*}

	\subsection{Excitations beyond linear analysis}
	We discuss nonlinear effects not captured under the linear analysis by taking nonlinear terms on the hydrodynamic variables. Here, we first focus on $f_z$ and $n_{zz}$ which are never resonated under the linear analysis.

	We rewrite the constraints~\eqref{bc:nf} and \eqref{bc:det} in terms of the fluctuations introduced in Eqs.~\eqref{eq:fluct1}--\eqref{eq:fluct4} as
\begin{align}
	\label{bc:del3}\delta f_x\delta n_{xy}+\delta f_y\delta n_{yz}+\delta f_z(\delta n_{zz}-1)=0,
	\end{align}
	\begin{align}
	\label{bc:det1}\left|\begin{array}{ccc}
	1+\delta n_{xx}&\delta n_{xy}&\delta n_{xz}\\
	\delta n_{xy}&1+\delta n_{yy}&\delta n_{yz}\\
	\delta n_{xz}&\delta n_{yz}&\delta n_{zz}
	\end{array}\right|=\frac{1}{4}\left[(\delta f_x)^2+(\delta f_y)^2+(\delta f_z)^2\right].
	\end{align}
	Note that the resonant variables $\delta f_x, \delta f_y, \delta n_{xz},$ and $\delta n_{yz}$ grow in time at resonance, and thus we expect that they are dominant in Eqs.~\eqref{bc:del3} and \eqref{bc:det1}. Taking the resonant (nonresonant) variables up to the second (first) order, we obtain
	\begin{align}
	\label{eq:del_fz}\delta f_z&\simeq\delta f_x\delta n_{xz}+\delta f_y\delta n_{yz}, \\\label{eq:del_nzz}\delta n_{zz}&\simeq\frac{1}{4}\left[ (\delta f_x)^2+(\delta f_y)^2 \right]+(\delta n_{xz})^2+(\delta n_{yz})^2.
	\end{align}
This means that $f_z$ and $n_{zz}$ can increase owing to the products of the two resonant variables. Next, we apply the Fourier transformation to Eqs.~\eqref{eq:del_fz} and \eqref{eq:del_nzz}, and obtain
\begin{align}
	\nonumber\delta\tilde{f}_z(k)\simeq\frac{1}{L_x}\sum_{k'}&\Big( \delta\tilde{f}_x(k')\delta\tilde{n}_{xz}(k-k')
\\ &+\delta\tilde{f}_y(k')\delta\tilde{n}_{yz}(k-k') \Big)\label{eq:kdel_fz},
	\\\nonumber\delta\tilde{n}_{zz}(k)\simeq\frac{1}{L_x}\sum_{k'}&\Big( \frac{1}{4}\delta\tilde{f}_x(k')\delta\tilde{f}_{x}(k-k')
\\ \nonumber&+\frac{1}{4}\delta\tilde{f}_y(k')\delta\tilde{f}_{y}(k-k')
\\ \nonumber&+\delta\tilde{n}_{xz}(k')\delta\tilde{n}_{xz}(k-k')
\\ &+\delta\tilde{n}_{yz}(k')\delta\tilde{n}_{yz}(k-k')\Big)\label{eq:kdel_nzz},
\end{align}
from which we find that $\delta\tilde{f}_z$ and $\delta\tilde{n}_{zz}$ have peaks at $k=0$ and $\pm2k_\mathrm{res}$ because the resonant variables grow around $k=\pm k_\mathrm{res}$. Note that $\tilde{f}_z$ is almost zero at $k=0$ because of the conservation law of the total longitudinal magnetization $\int f_z(\bm{r},t)d^dr$ which is a small value coming from the initial noise.

    The spectra of other nonresonant variables $\tilde{n}_{xx}, \tilde{n}_{yy}$, and $\tilde{n}_{xy}$ can be understood from directly approximating Eq.~\eqref{eq:SHE3} in the same manner as the above:
    \begin{align}
        \frac{\partial}{\partial t}\delta n_{xx}&\simeq\frac{2c_1\bar{\rho}}{\hbar}\delta f_y\delta n_{xz},
        \\\frac{\partial}{\partial t}\delta n_{yy}&\simeq\frac{2c_1\bar{\rho}}{\hbar}\delta f_x\delta n_{yz},
        \\\frac{\partial}{\partial t}\delta n_{xy}&\simeq\frac{c_1\bar{\rho}}{\hbar}(\delta f_y\delta n_{xz}-\delta f_x\delta n_{yz}), \label{B_end}
    \end{align}
    which implies that these variables grow around $k=0$ and $\pm2k_\mathrm{res}$. 

	We systematically investigate the peak structures of all the spin and nematic variables. 
	Our numerical calculations in Fig.~\ref{FIG:k9} demonstrate that the resonant variables ($\tilde{f}_x,\tilde{f}_y,\tilde{n}_{xz}$, and $\tilde{n}_{yz}$) have peaks around odd multiples of $k_\mathrm{res}$ and the others ($\tilde{f_z}, \tilde{n}_{xx}, \tilde{n}_{yy}, \tilde{n}_{zz}$, and $ \tilde{n}_{xy}$) have peaks around zero and even multiples of $k_\mathrm{res}$ in the early stage of the nonlinear dynamics. This is consistent with the results predicted by Eqs.~\eqref{eq:del_fz}-\eqref{B_end}. These spectra of the numerical results are similar to the typical pumped spectra seen in classical fluid \cite{Kolmogorov}. By taking the fluctuation terms to higher orders, we expect that all the spectra in the early stage in the figure are explained. 
	
	Using these results, we also understand the time evolution of $\rho_{\pm1}$, which is related to $n_{zz}$ and $f_z$ via
	\begin{align}
	\label{eq:rhopm1}\rho_{\pm1}=\rho\frac{n_{zz}\pm f_z}{2}.
	\end{align}
    Thus, $\tilde{\rho}_{\pm1}$ grows around zero and even multiples of $k_\mathrm{res}$. 
    This means that the increase in the particle number of the $m=\pm1$ component, which characterize the onset of the Shapiro resonances in the previous experimental works~\cite{Hoang2016,Shapiro1}, is attributed to the nonlinear effect.

	\begin{table*}[t]
		\centering
		\begin{tabular}{|c c r||l|l|l|l|l|l|l|} \hline
			Atomic species & dim. &Ref.& $c_1 \rho(\bm 0)/h$~[Hz] &$q_0/|c_1\rho(\bm 0)|$&$q_\mathrm{osc}/q_0$&$2\pi/k_\mathrm{res}$[$\mathrm{\mu m}$]&$\Omega/(2\pi)=2E_{k_\mathrm{res}}/h$~[Hz]& $\Delta\Omega/(2\pi)$~[Hz] & $\tau_\mathrm{FL}$~[sec] \\ \hline \hline
			$^{87}$Rb&1&\cite{Rb_Hirano} & $(-12.4)_\mathrm{TF}$ &$2.8$&$0.48$&5.3&$2.2\times10^2$& 2.5 & 0.25 \\ \hline
			$^{23}$Na&2&\cite{Na_Shin} & $(34.0)_\mathrm{TF}$ &$4.0$&0.3&(33, 23)&$3.7\times10^2$& 12 & 0.053 \\ \hline
			$^{23}$Na&1&\cite{Na_Raman} & $1.20\times10^2$ &1.13&0.1&68&$4.0\times10^2$& 11 & 0.059 \\ \hline
			$^{23}$Na&1&\cite{Na_Dalibard} & $(8.3)_\mathrm{TF}$ &$16$&0.9&29&$3\times10^2$& 9 & 0.07 \\ \hline
			$^7$Li&2&\cite{Li_Choi} & $-160$ &$1.9$&0.030&(15,12)&$4.8\times10^2$& 9.6 & 0.066 \\ \hline
		\end{tabular}
		\caption{Evaluation of the parameters according to the experiments in Refs.~\cite{Rb_Hirano,Na_Shin,Na_Raman,Na_Dalibard,Li_Choi} for the $J=1$ Shapiro resonance. In column $c_1\rho(\bm 0)$, the values marked with TF are calculated using the TF approximation, and the others are the values specified in the literature. In the evaluation of $\Omega, \Delta\Omega$, and $\tau_\mathrm{FL}$, we use the effective spin-dependent interaction energy $c_1'\bar{\rho}'$ instead of $c_1\bar{\rho}$, where $c_1'\bar{\rho}'$ is related to $c_1 \rho(\bm  0)$ via Eqs.~\eqref{eq:quasi1D_c1rho} and \eqref{eq:quasi2D_c1rho}.
		\label{table1}
		}
	\end{table*}
	
	\section{Discussion\label{Sec:discussion}}
	We discuss experimental possibility for observing spin-wave excitations due to the Shapiro resonance on the basis of our linear analysis of Sec.~\ref{sec:linear}. As shown in Sec.~\ref{subsec:FDMA}, the $J=1$ mode has a larger FL exponent than those with higher $J$. This means that the non-uniform Shapiro resonance with $J=1$ is experimentally more accessible. We, therefore,  address the resonance with $J=1$ by using realistic experimental parameters in what follows.

    To consider the experimental possibility, we introduce two important quantities: One is an inverse of the maximum FL exponent $\tau_\textrm{FL} \equiv (\lambda_{\rm FL}|_\mathrm{max})^{-1}$, and the other is the width $\Delta\Omega$ for the resonant frequency. Here, $\tau_\textrm{FL}$ is the characteristic time scale for the growth of spin waves, thus should be sufficiently longer than the time resolution of the experiments and shorter than a lifetime of a BEC. On the other hand, $\Delta \Omega $ should be narrow enough to specify the wavenumber of the resonant spin waves, but not too narrow so that it is easy to adjust in experiments.
    Here, we show two constraints for $\tau_\mathrm{FL}$ and $\Delta\Omega$ derived from Eqs.~\eqref{lambFL1} and \eqref{ieq:1}:
    \begin{align}
	   &\tau_\mathrm{FL} = \frac{2E_k\hbar}{q_\mathrm{osc}|c_1|\bar{\rho}} \geq \frac{\hbar}{|c_1|\bar{\rho}}, 
	   \label{dis1}\\
	    &\tau_\mathrm{FL} \Delta\Omega =4.
	\label{dis2}
	\end{align}
	The inequality~\eqref{dis1} indicates that the lower bound of $\tau_\mathrm{FL}$ is the characteristic time scale of spin dynamics, which is a few to a few tens of milliseconds in typical experiments.
	The constraint~\eqref{dis2} is a trade-off relation between $\Delta\Omega$ and $\tau_\mathrm{FL}$. We have used these constraints to choose the experimental parameters in the following discussion. 

    Using the obtained analytical results of Eqs.~\eqref{lambFL1} and~\eqref{ieq:1}, we evaluate $\tau_\textrm{FL}$ at $\hbar\Omega=2E_{k_\mathrm{res}}$ and $\Delta\Omega$ for the parameters in the experiments~\cite{Rb_Hirano,Na_Shin,Na_Raman,Na_Dalibard,Li_Choi}. We fix $q_0$ to be the QZ energy at the magnetic field of 700~mG and choose $q_\mathrm{osc}$ such that the polar state becomes stable for off-resonant $\Omega$. The values of the hyperfine splitting energy and the interaction strengths $c_0$ and $c_1$ are given in Refs.~\cite{YK2012,Na_scatlength,Rb_scatlength,Li_Choi}. As for the resonant wavenumber $\bm k_\mathrm{res}$, which determines $\hbar\Omega=2E_{k_\mathrm{res}}$, we choose $k_\mathrm{res}=10\pi/R_x$  for a quasi-1D BEC and $\bm k_\mathrm{res}=(10\pi/R_x,10\pi/R_y)$ for a quasi-two-dimensional (quasi-2D) BEC, where $R_x$ and $R_y$ are the largest and the second-largest Thomas-Fermi (TF) radii, respectively.
    Note that, in low-dimensional BECs, the spin interaction energy $c_1\bar{\rho}$ in Eqs.~\eqref{lambFL1} and~\eqref{ieq:1} and inequalities~\eqref{dis1} and \eqref{dis2} is replaced with the effective one, $c_1'\bar{\rho}'$, as derived in Appendix~\ref{app6_c1rho}. Here, $c_1'\bar{\rho}'$ is related to the peak density $\rho(\bm 0)$ of the 3D distribution via Eqs.~\eqref{eq:quasi1D_c1rho} and \eqref{eq:quasi2D_c1rho} for quasi-1D and quasi-2D BECs, respectively.
    
    We summarize the estimated values of $\tau_\mathrm{FL}$ and $\Delta\Omega$ in Table~\ref{table1}, together with the values of $c_1\rho(\bm 0),\ q_0,\ q_\mathrm{osc}, k_\mathrm{res}$, and $\Omega$. The obtained sets of values can be experimentally accessible.

	\section{Conclusion\label{Sec:conclusion}}
	In this work, considering the polar state in the uniform spin-1 BECs, we theoretically investigated the Shapiro resonance driven by a periodic forcing of the QZ term.
	Unlike the previous works which discuss the Shapiro resonance in a strongly confined BEC without spatial degrees of freedom~\cite{Hoang2016,Shapiro1}, we took into account the spatial dependence of the condensate and investigated the growth of spin-wave excitations.

	First, we studied the Shapiro resonance within the linear analysis by employing the spin hydrodynamic equations equivalent to the GPE. Applying the Floquet's theorem to the equations, we analytically obtained the real parts of the Floquet exponents, i.e., the FL exponents, featuring the growth rate of the resonant spin and nematic variables by using the two kinds of approximations: the finite-dimensional matrix approximation and the degenerate perturbation approximation. From these results, we identified the resonant conditions and found the spin-wave excitation with finite wave-numbers, which cannot be described by the single-mode approximation.

	Second, we numerically solved the spin-1 GPE and study the validity of the linear analysis and the nonlinear dynamics in the late stage. In the $k$-space time evolution of the hydrodynamic variables, we confirmed that the resonant variables are excited in the early stage of the dynamics as expected in the linear analysis. However, as time goes by, we found the emergence of spin-wave excitations at wavenumbers of integer multiples of the resonant one, which cannot be predicted by our linear analysis. We explained this nonlinear effect by using the constraints on the hydrodynamic variables and the spin hydrodynamic equations. Continuing to drive the QZ term, we investigated the long-time dynamics and observed that, in the AFM spinor BEC, the finer spin distributions emerged compared with the FM spinor BEC. 
	
	In the final section, we discussed the experimental possibilities for observing spin-wave excitations due to the Shapiro resonance on the basis of our linear analysis. Using the parameters used in the previous experiments~\cite{Rb_Hirano,Na_Shin,Na_Raman,Na_Dalibard,Li_Choi}, we showed that a driving QZ field at a frequency in the order of 100 Hz under a bias field of 700 mG can induce the growth of spin waves at the experimentally accessible length and time scales. Our results give an experimental procedure to excite spin waves of a specific wavenumber selectively, which would be useful for future studies of spin dynamics.

\begin{acknowledgments}
We would like to thank X. Chai, D. Lao, and C. Raman for fruitful discussions.
This work was supported by JST-CREST (Grant No. JPMJCR16F2), JSPS KAKENHI (Grant Nos. JP18K03538, JP19H01824,  JP19K14628, and JP20H01843), Toyota Riken Scholar, Foundation of Kinoshita Memorial Enterprise, and the Program for Fostering Researchers for the Next Generation (IAR, Nagoya University) and Building of Consortia for the Development of Human Resources in Science and Technology (MEXT). 
\end{acknowledgments}

	\appendix

	\section{\label{app2_eqs}Madelung form of the spinor GPE}
	In the main text, we use the spin hydrodynamic equations, but the previous studies deal with the Madelung form of the spinor GPE to investigate the Shapiro resonance under the single-mode approximation. Here, we explain the latter form and its consequence. 

	Substituting $\psi_m=\sqrt{\rho_m(\bm{r},t)}e^{-i\theta_m(\bm{r},t)}$ into the GPE before eliminating the linear Zeeman term $p$, we obtain the following equations from the real and imaginary part of the GPE:
	\begin{align}
		\frac{\partial}{\partial t}\rho_m-&\nabla\left[\rho_m\frac{\hbar}{M}\nabla\theta_m\right]=(6m^2-4)\frac{c_1}{\hbar}\rho_0\sqrt{\rho_1\rho_{-1}}\sin(\Delta\theta), \label{eq:Madelung_rho}
		\\\nonumber\hbar\frac{\partial}{\partial t}\theta_m= &\frac{\hbar^2}{2M}\left[-\frac{\nabla^2\sqrt{\rho_m}}{\sqrt{\rho_m}}+(\nabla\theta_m)^2\right]
		\\&-pm+qm^2+c_0\rho+G_m, \label{eq:Madelung_theta}
	\end{align}
	where $\Delta\theta=\theta_1+\theta_{-1}-2\theta_0$ and
	\begin{eqnarray*}
	G_m=\left\{
	\begin{array}{l}
	c_1[ \displaystyle{\rho_1+\rho_0-\rho_{-1}+\rho_0\sqrt{\frac{\rho_{-1}}{\rho_1}}\cos(\Delta\theta) }]\ (m=1);
	\\c_1[ \displaystyle{\rho_1+\rho_{-1}+2\sqrt{\rho_1\rho_{-1}}\cos(\Delta\theta) }]\ (m=0);
	\\c_1[ \displaystyle{-\rho_1+\rho_0+\rho_{-1}+\rho_0\sqrt{\frac{\rho_1}{\rho_{-1}}}\cos(\Delta\theta) }]\ (m=-1).
	\end{array}
	\right.
	\end{eqnarray*}
	
	These equations of motion elucidate that the phase difference $\Delta \theta$ induces changes in the number fraction in each magnetic sublevel. This mechanism is similar to the Josephson effect. Since the left-hand side of Eq.~\eqref{eq:Madelung_rho} includes the number densities of all components, the particle flow between spin components vanishes at the place where at least one of the densities $\rho_{0,\pm1}$ becomes zero. One can also see from the Madelung form that a spatially uniform linear Zeeman term $p$ does not affect the dynamics of $\rho_m$.

    The equations of motion~\eqref{eq:Madelung_rho} and \eqref{eq:Madelung_theta} certify the following identity:
	\begin{align}
		\label{eq:popu_N}\frac{dN_1}{dt}=\frac{dN_{-1}}{dt}=-2\frac{dN_0}{dt},
	\end{align}
    where $N_m\equiv\int\rho_md^dr$.
	This is due to the conservation of the longitudinal magnetization, $\int F_z d^d r$. Therefore, no resonance occurs when we start from a fully polarized state along $+z$ or $-z$ direction.

	\section{\label{app3_Q}Symmetry property of the FL exponents of Eq.~$(\ref{eq:S})$}
    We prove that Eq.~\eqref{eq:S} gives the same set of the FL exponents independently of the sign on the right-hand side. Below, we refer to Eq.~\eqref{eq:S} with plus and minus signs as Eqs.~(\ref{eq:S}$+$) and (\ref{eq:S}$-$), respectively, and denote their Floquet exponent as $\lambda_{\mathrm{F}+}$ and $\lambda_{\mathrm{F}-}$.
    The proof is two-step: (i) We first derive the relation $\mathrm{Re}[\lambda_{\mathrm{F}-}]=-\mathrm{Re}[\lambda_{\mathrm{F}+}]$;
    (ii) We then show that the Floquet exponents $\lambda_{\mathrm{F}\pm}$ and $-\lambda_{\mathrm{F}\pm}^*$ always appear in a pair, which means positive and negative FL exponents appear in a pair.
    Thus, Eqs.~(\ref{eq:S}$+$) and (\ref{eq:S}$-$) have the same set of FL exponents, and hence we solve only Eq.~\eqref{eq:EigenEq}, which corresponds to Eq.~(\ref{eq:S}$+$), in the main text.
    
	\textit{Step (i).-} Suppose that $\bm{S}_+(t)\equiv (S_{A+}, S_{B+})^\mathrm{T}$ is a solution of Eq.~(\ref{eq:S}$+$). By replacing  $t$ with $-t+\pi/\Omega$ in Eq.~(\ref{eq:S}$+$), we obtain
	\begin{align}
		\frac{d}{dt}\bm{S}_+\left(-t+\frac{\pi}{\Omega}\right)=-\left( \hat{F}+\frac{q_\mathrm{osc}}{E_k}\sin(\Omega t)\hat{G} \right)\bm{S}_+\left(-t+\frac{\pi}{\Omega}\right),
	\end{align}
	which indicates that
	\begin{align}
		\label{eq:S+S-}\bm{S}_-(t)= \bm{S}_+\left(-t+\frac{\pi}{\Omega}\right)
	\end{align}
	is a solution of Eq.~(\ref{eq:S}$-$). According to the Floquet's theorem, the solutions $\bm S_\pm(t)$ can be rewritten as $\bm{S}_\pm(t)=e^{\lambda_\mathrm{F\pm}t}\bm{p}_\pm(t)$ with a periodic function $\bm{p}_\pm(t)=\bm{p}_\pm(t+T)$. Using these forms, Eq.~(\ref{eq:S+S-}) is rewritten as
	\begin{align}
		e^{\lambda_\mathrm
		{F-}t}\bm{p}_-(t)=e^{\lambda_\mathrm{F+}(-t+\pi/\Omega)}\bm{p}_+\left(-t+\frac{\pi}{\Omega}\right).
	\end{align}
	 Comparing the $t$-dependencies of both sides, we obtain $\mathrm{Re}[\lambda_\mathrm{F+}]=-\mathrm{Re}[\lambda_\mathrm{F-}]$ and $\mathrm{Im}[\lambda_\mathrm{F+}]=-\mathrm{Im}[\lambda_\mathrm{F-}]$ mod $\Omega$.
	 
	\textit{Step (ii).-}
	Suppose that $\hat{Q}$ has an eigenvalue $\varepsilon$.
	Since $\hat{Q}$ is not an Hermitian matrix, it has right and left eigenstates:
	\begin{subequations}
	\begin{align}
	    \hat{Q}|C\rangle &= \varepsilon |C\rangle,\\
	    \langle \tilde{C}|\hat{Q} &= \langle \tilde{C} | \varepsilon.
	\end{align}
	\end{subequations}
	By taking the Hermitian conjugate of the second equation, we obtain
	\begin{align}
	    \hat{Q}^\dagger|\tilde{C}\rangle = \varepsilon^* |\tilde{C}\rangle.
	    \label{eq:left_eigen_vector}
	\end{align}
	Here, we use the psudo-Hermiticity of $\hat{Q}$:
	As we have explained in Sec.~\ref{sec:resonant_conditions}, $\hat{Q}$ is a psudo-Hermitian matrix and satisfies
	\begin{align}
	     \hat{Q}^\dagger  = \eta^\dagger\hat{Q}\eta,
    \label{eq:psudoH}
	\end{align}
	with $\eta=\mathrm{Diag}[\cdots,1,-1,1,-1,\cdots]$.
	By substituting Eq.~\eqref{eq:psudoH} into Eq.~\eqref{eq:left_eigen_vector} and multiplying by $\eta$ from the left, we obtain
	\begin{align}
	    \hat{Q}\eta |\tilde{C}\rangle = \varepsilon^* \eta|\tilde{C}\rangle.
	\end{align}
	That is, $\varepsilon^*$ is also an eigenvalue of $\hat{Q}$, and the corresponding eigenstate is given by $\eta|\tilde{C}\rangle$.
	Since $(i\lambda_\mathrm{F})$ is an eigenvalue of $\hat{Q}$ [see Eq.~\eqref{eq:EigenEq}], $(i\lambda_\mathrm{F})^*=i(-\lambda_\mathrm{F}^*)$ is also an eigenvalue of $\hat{Q}$. It follows that if there is a nonzero FL exponent $\lambda_\mathrm{FL}=\mathrm{Re}[\lambda_\mathrm{F}]$, there is always another nonzero FL exponent $-\lambda_\mathrm{FL}=\mathrm{Re}[-\lambda_\mathrm{F}^*]$.
		
	\section{\label{app4_b} Validity of the finite-dimensional matrix approximation and the perturbation approximation}
	The off-diagonal elements of the matrix $\hat{Q}$ in Eq.~\eqref{matrix:Q} is proved to be less than the half of the diagonal ones under the polar regime given by Eq.~(\ref{ieq:polar}):
	\begin{align*}
	\left|\frac{b(\hat{G})_{l,m}}{E_k}\right|&<\frac{q_\mathrm{osc}(\epsilon_k+q_0+c_1\bar{\rho})}{2E_k^2}
	\\&=\frac{q_\mathrm{osc}(\epsilon_k+q_0+c_1\bar{\rho})}{2(\epsilon_k+q_0)(\epsilon_k+q_0+2c_1\bar{\rho})}
	\\&<\left\{\begin{array}{ll}
	\displaystyle{ \frac{q_\mathrm{osc}(\epsilon_k+q_\mathrm{osc}+|c_1|\bar{\rho})}{2(\epsilon_k+q_\mathrm{osc})(\epsilon_k+q_\mathrm{osc}+2|c_1|\bar{\rho})} }& (c_1<0)
	\\\displaystyle{ \frac{q_\mathrm{osc}}{2(\epsilon_k+q_0)} }& (c_1>0)
	\end{array}\right.
	\\&<\frac{1}{2}.
	\end{align*}
	Here, we use the condition~\eqref{ieq:polar} from the second to the third line. This confirms the validity of the finite-dimensional matrix approximation and the perturbation approximation in this paper.
	
	\section{\label{app5_J=2}Finite-dimensional matrix approximation solution in $\bm{J=2}$}
    In the finite-dimensional matrix approximation, the eigenvalue equation~\eqref{eq:EigenEq_J=2} of the $8\times 8$ matrix for the $J=2$ resonance obviously has a solution around Eq.~\eqref{intsec}. Therefore, we expand the solution as $\Omega=E_k/\hbar+\delta\Omega$ and   $i\lambda_\mathrm{F}=jE_k/\hbar+i\delta \lambda_\mathrm{F}$, which corresponds to $(j_1,j_2)=(j-1,j+1)$, and take up to the second order with respect to $\delta \lambda_\mathrm{F}$ and $\delta\Omega$. Then, the equation becomes quadratic and the solution is given by
	\begin{align}
		\label{eq:FDMAJ=2}i\delta \lambda_\mathrm{F}=-j\delta\Omega\pm\sqrt{\frac{Y}{X}}
	\end{align}
	where
	\begin{align*}
		X&=36\alpha^8+288\alpha^7\gamma+14\alpha\beta^4\gamma^3+3\beta^4\gamma^4
		\\&+36\alpha^6(\beta^2+24\gamma^2)+72\alpha^5(3\beta^2\gamma+16\gamma^3)
		\\&+\alpha^4(9\beta^4+448\beta^2\gamma^2+576\gamma^4)
		\\&+4\alpha^3(9\beta^4\gamma+88\beta^2\gamma^3)+\alpha^2(43\beta^4\gamma^2+64\beta^2\gamma^4),
		\\Y&=(\delta\Omega)^2[36\alpha^8+288\alpha^4\gamma+72\alpha^5\gamma(3\beta^2+16\gamma^2)
		\\&36\alpha^6(\beta^2+24\gamma^2)-8\beta^4\gamma^4-64\alpha\beta^4\gamma^3+640\alpha^2\beta^2\gamma^4
		\\&+4\alpha^2\beta^4\gamma^2+576\alpha^4\gamma^4+592\beta^2\gamma^2+9\alpha^4\beta^4-928\alpha^3\beta^2\gamma^3
		\\&+18\alpha^3\beta^4\gamma]
		\\&+(\delta\Omega) E_k[-16\beta^4\gamma^4-72\alpha\beta^4\gamma^3+96\alpha^2\beta^2\gamma^4-36\alpha^2\beta^4\gamma^2
		\\&+24\alpha^4\beta^2\gamma^2+96\alpha^3\beta^2\gamma^3]
		\\&-9\alpha^4\beta^4\gamma^2-36\alpha^3\beta^4\gamma^3-41\alpha^2\beta^4\gamma^4-10\alpha\beta^4\gamma^4.
	\end{align*}
	Here, we define $\alpha=\epsilon_k+q_0, \beta=q_\mathrm{osc},$ and $\gamma=c_1\bar{\rho}$. The above results were obtained using Mathematica. In Fig.~\ref{FIG:lamNum}(b), the curves labeled with $\lambda_\textrm{FL}^\textrm{(ana)}$ represents the positive imaginary part of $\delta \lambda_\textrm{F}$ in Eq.~\eqref{eq:FDMAJ=2}, showing good agreement with the numerically obtained one. Note that although we have calculated for $(j_1,j_2)=(j-1,j+1)$, the imaginary part of Eq.~\eqref{eq:FDMAJ=2} does not depend on $j$, suggesting that the FL exponent, i.e., $\textrm{Im}(\lambda_\textrm{F})$, is the same for all combinations of $(j_1,j_2)$ satisfying $j_2-j_1=2$.
	We can also confirm that the FL exponent at the resonance point $\hbar\Omega=E_k$ coincides with Eq.~\eqref{eq:lam2} by substituting $\hbar\Omega=E_k$ into Eq.~\eqref{eq:FDMAJ=2} and expanding it with respect to $b$ up to the second order.
	
	\section{Effective spin-dependent interaction energy in a low-dimensional BEC\label{app6_c1rho}}
    Although our theoretical analysis in the main text deals with a uniform system, many experiments prepare low-dimensional BECs strongly confined in one or two directions. For such systems, we obtain the same results as in the main text by using the low-dimensional GPE, where the interaction energies are replaced with the ones averaged along with the directions of strong confinement. Below, we derive the effective spin-dependent interaction energy, $c_1' \bar{\rho}'$, for quasi-1D and quasi-2D BECs, following Ref.~\cite{chai2020magnetic}.
	
	We start from describing how to derive the 1D GPE. We suppose that the confinement along the $y$- and $z$-directions is strong enough such that the cloud size along these directions are smaller than the length scale of spin waves we are considering. In such a case, as in Ref.~\cite{chai2020magnetic}, we can factorise the macroscopic wave function as $\psi_m(x,y,z,t)=\Psi_m^{\rm 1D}(x,t)G^{\rm 1D}(y,z)$. Here, $G^{\rm 1D}(y,z)$ is given by the TF distribution:
	\begin{align}
        G^{\rm 1D}(y,z)=\sqrt{\frac{2}{\pi R_yR_z}\left( 1-\frac{y^2}{R_y^2}-\frac{z^2}{R_z^2} \right)}, 
    \end{align}
    where $R_y$ and $R_z$ are the TF radii, and we set the origin of the spatial coordinate to be the center of the harmonic potential. Then, integrating the 3D GPE with $y$ and $z$, we can derive the 1D GPE with the effective interaction coefficients $c_j'=4c_j/(3\pi R_yR_z) \ (j=0,1)$. 
    After the same calculation as in the text with the 1D GPE, we obtain Eqs.~\eqref{lambFL1} and~\eqref{ieq:1} where $c_1\bar{\rho}$ is replaced with $c'\bar{\rho}'$. Here, $\bar{\rho}'= |\Psi_0^{\rm 1D}(0,0)|^2$ is the 1D number density of the spatially uniform initial state in the polar state, which is related to the 3D initial number density at $\bm x=\bm 0$ via $\rho({\bm 0})=\bar{\rho}'[G^{\rm 1D}(0,0)]^2$.
    Thus, we can use Eqs.~\eqref{lambFL1} and~\eqref{ieq:1} by replacing $c_1\bar{\rho}$ with 
    \begin{align}
        c_1'\bar{\rho}'=\frac{2}{3}c_1\rho(\bm{0})~~~(\mathrm{1D}).
    \label{eq:quasi1D_c1rho}
    \end{align}
    
    For the case of a quasi-2D system with a strong confinement along the $z$ axis, we assume the factorization $\psi_m(\bm{r},t)=\Psi_m^{\rm 2D}(x,y,t)G^{\rm 2D}(z)$, where $G^{\rm 2D}(z)$ is given in the TF approximation by
    \begin{align}
        G^{\rm 2D}(z)=\sqrt{\frac{3}{4R_z}\left(1-\frac{z^2}{R_z^2}\right)}. 
    \end{align}
    Then, by integrating the 3D GPE with $z$, we obtain the 2D GPE with the effective interaction coefficients $c_j'=3/(5R_z)$ ($j=0,1$). 
    Thus, as in the case of 1D GPE, we obtain Eqs.~\eqref{lambFL1} and~\eqref{ieq:1} where $c_1\bar{\rho}$ is replaced with 
    \begin{align}   
        c_1'\bar{\rho}'=\frac{4}{5}c_1\rho(\bm{0})~~~(\mathrm{2D}).
    \label{eq:quasi2D_c1rho}
    \end{align}
    In the 2D case, $\bar{\rho}'=|\Psi_0^{\rm 2D}(0,0,0)|^2$ is the 2D number density of the spatially uniform initial state in the polar state, and we have used the relation $\rho({\bm 0})=\bar{\rho}'[G^{\rm 2D}(0)]^2$.

	\bibliography{reference}
\end{document}